\documentclass[aps,prd,preprint,tightenlines,groupedaddress,showpacs]{revtex4}
\usepackage{amssymb,amsmath,graphicx}

\usepackage{verbatim,color,ulem,comment}
\begin{document}
\title{ Quantum noise in the mirror-field system: A field theoretic approach}
\author{Jen-Tsung Hsiang}
\author{Tai-Hung Wu}
\author{Da-Shin Lee}
\email{dslee@mail.ndhu.edu.tw}

\affiliation{Department of Physics, National Dong-Hwa
University, Hua-lien, Taiwan, R. O. C.}
\author{ Sun-Kun King}
\affiliation{Academia Sinica, Institutes for Astronomy and Astrophysics, Taipei, Taiwan, R.O.C. }
\author{Chun-Hsien Wu}
\affiliation{Department of Physics, Soochow University, Taipei, Taiwan, R.O.C.}
\date{\today}
\begin{abstract}
We employ the field theoretic approach to study the quantum noise
problem in the mirror-field system, where a perfectly reflecting
mirror is illuminated by a single-mode coherent state of the
massless scalar field. The associated radiation pressure is
described by a surface integral of the stress-tensor of the field.
The read-out field is measured by a monople detector, form which the
effective distance between the detector and mirror can be obtained.
In the slow-motion limit of the mirror, we are able to identify
various sources of quantum noise that lead to uncertainty of the
read-out measurement. Since the mirror is driven by radiation
pressure, the sources of noise, other than the shot nose given by
the intrinsic fluctuations of the incident state, may also result
from random motion of mirror due to radiation pressure fluctuations
and from modified field fluctuations induced by the displacement of
the mirror. Correlation between different sources of noise can be
established as the consequence of interference between the incident
field and the reflected field out of the mirror in the read-out
measurement. The overall uncertainty is found to decrease (increase) due to the negative (positive)
correlation. In the case of negative correlation, the uncertainty
can be lowered than the value predicted by the standard quantum
limit. We also examine the validity of the particle number approach,
which is often used in quantum optics, and compared its results with those given by the field theoretical approach. Finally we
discuss the backreaction effects, induced by the radiation pressure,
that alter the dynamics of the mean displacement of the mirror, and
we argue this backreaction can be ignored for a slowly moving
mirror.
\end{abstract}

\pacs{03.70.+k, 05.40.-a, 12.20.Ds, 42.50.Dv, 42.50.Lc}

\maketitle
\baselineskip=18pt
\allowdisplaybreaks

\section{Introduction}
The measurement of a very weak force such as the
detection of gravitational wave demands a very high resolution
interferometer. The role of the interferometer is to transform small
variation of the relative spatial separation of the mirrors between
two arms into variation of the output photons. Quantum noise
presumably places a limitation on accuracy of the
measurement~\cite{CAT}.

In quantum optics, the discussion related to quantum noise of
radiation fields on the mirror has been mostly based on the
photon-number approach. It is argued by Caves~\cite{CA1,CA2} that
quantum noise in a laser interferometer may come from the quantum
nature of the light directly via photon number fluctuations (shot
noise) or indirectly via random motion of the mirror due to a
fluctuating force (radiation pressure fluctuations).  Minimizing the
total uncertainty from two sources of noise, with an assumption of
no correlation between them, may give the standard quantum limit
(SQL), when an input power of the light is appropriately tuned.
Additionally, Caves further suggests that injecting squeezed vacuum
state into the unused port of an interferometer should lead to the
SQL with a lower optimum input power~\cite{CA2}, as compared with
the standard interferometer~\cite{CA1}. This seminal work also shows
the manipulated quantum fluctuations of the electromagnetic fields
are responsible for both the radiation-pressure forces and the
fluctuations of the photon number. Then, an inquiry into the
assumption of no correlation between these two
noise sources arises. In particular, the presence of negative
correlations may lead to the uncertainly even below the SQL. To
establish their correlations requires an unified scheme that has
been proposed by Loudon~\cite{LO}. The judicious
use of the correlated squeezed state is proposed to establish the
above correlations with which to possibly push the sensitivity of
the interferometer beyond the SQL~\cite{JA}. The redesign of an
interferometer by adding some optical components to either
manipulate the read-outs or modify the dynamics of the test mass,
such as in the ``optical bar''~\cite{BAG} or the ``optical
spring''~\cite{BU} scheme, is also possibly to give non-zero
correlations in order to beat the SQL.

The coupling between the mirror and the radiation
field can be obtained by considering the mirror as a reflector that
imposes boundary conditions of the field on the surface of the
mirror. In 3+1 dimensions, field quantization is treated
perturbatively for a slowly moving, perfectly reflecting mirror by Vilenkin and
Ford~\cite{FOV}, with the boundary conditions of the field on the surface of the mirror at all times. They then
explore the backreaction dissipative effect on a moving mirror,
induced by vacuum fluctuations of the field. Later a corresponding
Langevin equation was derived by two of us~\cite{WU} with the method
of influence functional, in which the accompanying noise, manifested
from vacuum fluctuations is taken into account.
Thus, the effects of the backreaction dissipation
and its accompanying noise can be related by the underlying
fluctuation-dissipation relation. Unruh~\cite{UN} extended the
quantization scheme to the interferometer case where the field
propagates toward a mirror and then
acquires a time-dependent phase shift after reflecting off the
mirror. Because of the moving boundary, the phase shift in general
depends on the mirror's motion, but can be obtained by assuming
slow-motion of the mirror. Similar phase shift was also introduced
by Kimble {\it et al.}~\cite{KIM}, and it has been proposed to
improve performance of the laster interferometer, particularly,
in~\cite{BU}.

\begin{figure}
\centering
    \scalebox{1.20}{\includegraphics{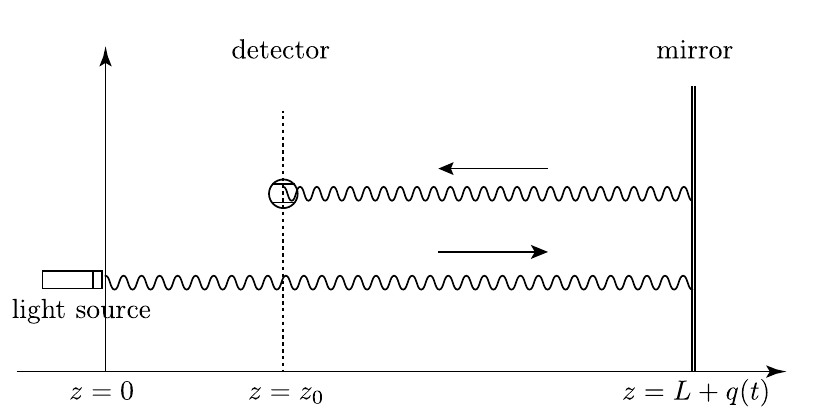}}
    \caption{Schematic diagram of the field-mirror system.}\label{Fi:dksl}
\end{figure}

Here we adopt the field quantization scheme similar to Unruh's
proposal to investigate the problem of quantum noise and its
associated correlation in a simpler configuration, shown in
Fig.~\ref{Fi:dksl}. The basic idea is to consider a free, perfectly
reflecting mirror. A coherent  state of the quantum scalar field is
normally shined on the mirror's surface, and exerts a pressure force
on the mirror. It drives the mirror into motion, which in turn leads
to variation of the reflected field. Then the read-out field is
measured by a standard monopole detector, which is placed somewhere
between the mirror and the field source. Since the radiation
pressure is described by a surface integral of the stress-tensor of
the scalar field, if the input quantum state is not an eigenstate of
the operator associated with the response function of the detector
or the stress-tensor, the resulting measurement will exhibit
fluctuations.

The goal of this paper is two-fold. First of all, we lay out a
field-theoretic approach by which the effective distance between the
detector and the mirror can be decoded from the read-out fields. In
this simpler system, the sources of the uncertainty in the read-outs
can be easily identified in the slow motion limit when the mirror is
driven by radiation pressure of the incident field. Thus reduction
in uncertainty of this effective distance is of essence to improve
the sensitivity of the interferometer. When the incident field is in
the single-mode coherent state, it will be shown that the particle
number approach, typically used in quantum optics, is equivalent to
this field theoretic approach in the late-time limit with a large
number of the scalar particles (photon). We will elaborate the
previously mentioned approximations later. In addition to shot noise
and radiation pressure fluctuations, a new source of noise comes
from modified field fluctuations. It is induced by the mean
displacement of the mirror as a result of the radiation pressure.
Its effect can be systematically examined in the field theoretic
approach. Secondly, the emphasis is put on how correlation among all
sources of quantum noise associated with the radiation field, can
emerge when the interference of the incoming field and the reflected
filed by the mirror are taken into account. We further examine the
effects of correlation on possible reduction of uncertainty in the
read-outs of the monopole detector. We show an example in which the
overall uncertainty can be suppressed to beat the standard quantum
limit.

Our presentation is organized as follows. In Sec.~\ref{sec1}, we
introduce the quantization scheme of a massless scalar field subject
to boundary conditions on a free, perfectly reflecting mirror when
it undergoes slow motion. The motion of the mirror arises from
radiation pressure of the scalar field. A monopole detector is
introduced to measure the readout field. We then identify the
sources of quantum noise from the fluctuations of the readouts in
Sec.~\ref{sec2} with an emphasis on establishing correction between
them. The effects of correlation are further studied in
Sec.~\ref{sec3} for possibility of reducing overall quantum noise.
In Sec.~\ref{sec4} the Langevin equation of the mirror, including
backreaction from the radiation field, is derived, and we show that
this effect is negligible for slow motion. Conclusions are drawn in
Sec~\ref{sec6}.

The Lorentz-Heaviside units and the convention $\hbar=c=1$ will be used unless otherwise mentioned. The signature of the metric is $\operatorname{diag}\{\eta_{\mu\nu}\}=(+1,-1,-1,-1)$.

\section{field quantization and detector theory}\label{sec1}
We consider a free mirror of perfect reflectance is illuminated by a
single-mode coherent laser, which propagates in a
direction normal to the mirror. The radiation pressure of the laser
will drive the mirror into motion, and the field, bounced off the
mirror, can be measured by a detector.
For simplicity and without loss of generalization, the radiation
field is represented by a massless scalar field $\phi$ to suppress
complexity caused by the vector nature of the electromagnetic field.
The mirror, with mass $m$, is placed at $z=L$. Its lateral
dimensions in the $x$, $y$ directions are large compared to
wavelength of the incoming state, and the cross sectional area is
$\mathcal{A}_{\parallel}$. We use $\mathbf{x}_{\parallel}$ to denote
directions normal to the $z$ axis. The incident field is a plane
wave propagating along the positive $z$ direction toward the mirror.
The boundary condition of the field on the mirror surface is then
given by
\begin{equation}
\phi (\mathbf{x}_{\parallel},L+q (t),t)=0 \, , \nonumber
\end{equation}
where $q(t)$ is the displacement of the mirror from its original position at $ z=L$ due to radiation pressure. The approximate solution to the field equation subject to this moving boundary condition in the region of $z<L$ is ~\cite{UN}
\begin{equation}
    \phi =\phi^+ +\phi^- \, , \nonumber
\end{equation}
where the positive-frequency solution $\phi^+$ is given by
\begin{equation}
    \phi^{+}(\mathbf{x},t)=\int'\!\!\frac{d^3k}{(2\pi)^{3/2}}\frac{1}{\sqrt{2\omega}}\;a_{k}\,\mathcal{U}_{\mathbf{k}}(\mathbf{x},t)\,,
\end{equation}
with
\begin{equation}
    \mathcal{U}_{\mathbf{k}}(\mathbf{x},t)=e^{i
\mathbf{k}_{\parallel} \cdot \mathbf{x}_{\parallel}-i\omega t}\left[e^{ik_zz}-e^{-ik_z(\,z-2L-2q(t_{R})\,)}\right] \, .\label{mode}
\end{equation}
The negative-frequency solution $\phi^-$ is the Hermitian conjugate
of the positive-frequency solution $\phi^{+}$, that is, $\phi^-=(\phi^+)^{\dagger}$. The
prime over the integrals of $\mathbf{k}$ means that the modes with
$k_z <0$ are excluded because these modes propagate in the negative
$z$ direction and will not interact with the mirror in the $z<L$
region. Here we have assumed that the mirror undergoes slow motion
($\dot{q} \ll 1$). Since the incoming wave is a right-moving plane
wave along the $z$ direction, the left-moving reflected wave will
receive a time-dependent phase shift due to the motion of the
mirror. The phase shift of the reflected field at time $t$ and
position $z$ depends on the mirror's position  at an earlier time
$t_{R}= t-\lvert L+q(t_{R})-z\rvert$ if we take finite lapse of
propagation into consideration. When the displacement is small, we
may take the retarded time $t_{R}$ as $t_{R}= t-\lvert L-z\rvert$ in
the leading order approximation.

Due to small displacement approximation, the force acting on the mirror by the fields can be approximated by the area integral of the $zz$ component of the energy-momentum stress tensor of the scalar field, evaluated at the mirror's original position $z=L$,
\begin{equation}\label{E:force}
    F_0(t)=\int\!d\mathcal{A}_{\parallel}\;T_{zz}(\mathbf{x}_{\parallel},L,t)\,,
\end{equation}
where the $zz$ component of the energy-momentum stress tensor takes the form
\begin{equation}
    T_{zz}(\mathbf{x}_{\parallel},z,t)= \frac{1}{2} \left[\bigl(\partial_t\phi\bigr)^2
+\bigl(\partial_z\phi\bigr)^2-\bigl(\partial_{\mathbf{x}_{\parallel}}\phi\bigr)^2 \right]\,.
\end{equation}
Hereafter we will use $d\mathcal{A}_{\parallel}$ to denote the area
element normal to the $z$ direction. Thus, if the mirror is at rest
at $z=L$ for $t<0$, then the displacement operator due to the action
of radiation pressure of the incident state is
formally given by
\begin{equation} \label{q_F}
    q(t) =\frac{1}{m}\int_{0}^{t}\!ds\!\int_{0}^{s}\!ds'\!\int\!d\mathcal{A}_{\parallel}\;T_{zz}({\mathbf{x}}_{\parallel},L,s')\,.
\end{equation}
Since the energy-momentum stress tensor is quadratic in the field operator $\phi$, the displacement operator is not well-defined. A proper renormalization of mirror's position operator can be done by absorbing the vacuum expectation value of the displacement operator into the initial position $L$, namely,
\begin{equation*}
    q(t)+L=q(t)-\langle q(t)\rangle_0+\bigl(L+\langle q(t)\rangle_0\bigr)=q_{r}(t)+L_{r}\,.
\end{equation*}
This prescription is physically sensible since
when the mirror is placed at its initial position, the vacuum
fluctuations of the field have already interacted with the mirror
before the incoming state impinges on it. Thus, the renormalized
displacement operator $q_{r}(t)$ now takes the form
\begin{align*}
    q_{r}(t)&=\frac{1}{m}\int_0^t\!d{s}\!\int_0^{s}\!ds'\!\int\!d\mathcal{A}_{\parallel}\;\Bigl\{T_{zz}({\mathbf{x}}_{\parallel},L,s')-\langle T_{zz}({\mathbf{x}}_{\parallel},L,s')\rangle_{0}\Bigr\}\,.
\end{align*}
Here the backreactions of the field  is ignored. This backreaction
can be incorporated in the Langevin equation, Eq.~\eqref{E:EoM}, and
is found negligible for a slowly moving mirror. To avoid cluttering
the notion, we will suppress the subscript $r$ of the renormalized
quantities hereafter.

To measure the outcomes of the scalar field, we employ a monopole
detector. We assume that the detection processes is based on the
stimulated transition of the detector due to coupling between the
scalar field and the monopole moment $m(t)$ of the detector,
\begin{equation}
    \int^{t}\!ds\,m(s)\,\phi(\mathbf{x},s)\, .
\end{equation}
In the standard detector theory, the dynamics of the detector is governed by some time-independent Hamiltonian $H_{D}$. Assume that the detector is a two-level system with eigen-energies $E_1$, $E_2$ such that $H_{D}\lvert E_{1,2}\rangle=E_{1,2}\lvert E_{1,2}\rangle$ and $E_1 < E_2$. In the interaction picture the monopole moment evolves in time as
\begin{equation}
    m(t) = e^{iH_{D}t}\,m(0)\,e^{-iH_{D}t}\,.
\end{equation}
Then by first-order perturbation theory, the transition rate between these two states of the detector is given by
\begin{equation}
    P (E_1\rightarrow E_2 )=\lvert\langle E_1\vert m(t)\vert E_2\rangle\rvert^2\times\Pi_{\phi}(E_2-E_1)\,.
\end{equation}
The response function $\Pi_{\phi}$ is defined by
\begin{align}
    \Pi_{\phi}(E)&=\int\!ds\!\int\!ds'\; e^{-iE(s-s')}\,\langle\phi^-(\mathbf{x},s)\phi^+ (\mathbf{x},s') \rangle_{\bar{\alpha}}\notag\\
    &=\delta(E-\bar{\omega})\int\!ds\;\langle\phi^-(\mathbf{x},s)\phi^+ (\mathbf{x},s)\rangle_{\bar{\alpha}}\,,\label{E:sdkhwk}
\end{align}
where we have assumed that the incident field is in a single-mode coherent state $\lvert\bar{\alpha}\rangle$ with frequency $\bar{\omega}$. Eq.~\eqref{E:sdkhwk} implies that we may construct an integral quantity by
\begin{equation*}
    I(z_0,t;L+q)=\int_0^t\!ds\!\int\!d\mathcal{A}_{\parallel}\;S(\mathbf{x}_{\parallel},z_0,s;L+q) \,,
\end{equation*}
with
\begin{equation} \label{I_q}
    S(x;L+q)=\phi^-(x)\,\phi^+(x)\,,\qquad\qquad\text{and $x=(t,\mathbf{x}_{\parallel},z_0)$}\,,
\end{equation}
where the detector is placed at $z=z_{0}$ and the position of the
mirror is located at $z=L+q(t)$. The quantity $I$ is a measure that
generically summarizes accumulated measurement
results over finite time duration. Integration over an area of the
mirror size illustrates the situation that if the size of the
detector is much greater than $2\pi/\bar{k}_{\parallel}$ of the
incoming coherent mode, then the measurements have been averaged
in the lateral spatial
direction. Then in terms of the mode functions of the scalar field,
$I(z_0,t;L+q) $ can be expressed as
\begin{equation}\label{I_L}
    I(z_0,t;L+q)=\int_0^t\!ds\!\int\!d\mathcal{A}_{\parallel}\!\int'\!\!\frac{d^3k}{(2\pi)^{\frac{3}{2}}}\!\int'\!\!\frac{d^3k'}{(2\pi)^{\frac{3}{2}}}\;\frac{1}{\sqrt{2\omega}}\frac{1}{\sqrt{2\omega'}}\;\mathcal{U}^*_{\mathbf{k}}(\mathbf{x}_{\parallel},z_0,s)\mathcal{U}_{\mathbf{k}'}(\mathbf{x}_{\parallel},z_0,s)\,a^{\dagger}_{\mathbf{k}}a_{\mathbf{k}'}^{\vphantom{\dagger}}\,.
\end{equation}
For the coupled mirror-field system, we notice
that they are at least two sources that will contribute to the
uncertainty associated with $I$. One directly results from quantum
fluctuations of the incoming field, but the other
indirectly comes from the stochastic motion of the
mirror, which in turn is the consequence of radiation-pressure
fluctuations of the field. In other words, the quantum fluctuations
of the incoming field will drive the mirror into stochastic motion,
and next the moving mirror will modify the reflected field in a
complicated way, depending on mirror's motion. If we use the a
detector to measure field variation at $z=z_{0}<L$, the detector
will see combined effects from the incoming and the reflected field.
It will be shown later that the interference between the
right-moving (incoming) and left-moving (outgoing) field leads to
the result that the phase of the expectation value of the operator
$I(z_0,t;L)$ depends on the effective distance between the mirror
and the detector. In addition, this interference effect is essential
to establish correlation between different sources of quantum noise.
This phenomenon gives a hope to reduce
overall quantum noise.

\section{quantum noise}\label{sec2}
Since the mirror is driven by the radiation pressure of a
single-mode coherent state with the frequency
$\bar{\omega}$, the assumption of slow motion of the mirror can be
realized by
\begin{equation}
    \bar{\omega}\,q(t) \ll 1 \,,\qquad\qquad q(t)/L\ll 1\,.
\end{equation}
This suggests that we may formally expand the field operators,
$\phi^{\pm} (x; q+L)$ and  $I(z_{0},t; q+L) $, in  a power series of
the the displacement operator $q(t)$. The operator ordering can be
quite ambiguous in this case. However, it can be shown that the
leading-order result is independent of operator ordering. Thus, for
example, up to the second order in $q$, we may just expand $I$ into
\begin{equation}\label{I_T_q2}
    I(z_0,t; q+L)= I(z_{0},t; L)+q(t_{\rm{ret}})\,\partial_L I (z_0 ,t;L) +\frac{1}{2}\,q^2 (t_{\rm{ret}}) \,\partial^2_L I (z_0,t;L)+\mathcal{O}(q^3)\,,
\end{equation}
irrespective of the operator ordering. Apparently, the first term
represents the measurement performed with respect to the mirror's
original position. The other terms can be
understood as corrections which result from the mirror's motion due to
radiation pressure.

The single-mode coherent state of the background field can be
obtained by applying the displacement operator $D(\bar{\alpha})$ on
the corresponding vacuum state,
\begin{equation}
    \lvert\bar{\alpha}\rangle=D(\bar{\alpha}) \lvert0\rangle\,,
\end{equation}
where the displacement operator $D(\bar{\alpha})$ is defined by
\begin{equation*}
    D(\bar{\alpha})= \exp\bigl[\bar{\alpha}\, a^{\dagger}_{\bar{\omega}} -\bar{\alpha}^{*}\, a^{\vphantom{\dagger}}_{\bar{\omega}}] \,.
\end{equation*}
Then, for a fixed position of the mirror, the signal is given by the expectation value
of $\langle\,I(z_0,t;L)\,\rangle_{\bar{\alpha}}$, which is obtained
from Eq.~\eqref{I_L} by setting $q=0$
\begin{equation}\label{I_exp}
    \langle \, I (z_0,t; L) \, \rangle_{\bar{\alpha}}  = \frac{\mathcal{A}_{\parallel}}{(2\pi)^{3}}\,\lvert\bar{\alpha}\rvert^2\,\frac{2}{\bar{\omega}}\,\sin^2\bar{\omega}(L-z_0)\times t\,.
\end{equation}
The result depends on the relative distance
between the detector plane and the mirror surface, which  arises
from the optical path difference between the right-moving (incoming)
field and the left-moving (outgoing) field. If a free
mirror, instead of a fixed one, is considered, the above result
will be modified to account for the additional effect due to the
mean displacement of the mirror. We further assume that the coherent
state under consideration has a sufficiently large particle density
$\lvert\bar{\alpha}\rvert^2$ so that in this state, the ratio of the
operator's fluctuations over its expectation value, which is
typically inversely proportional to $\lvert\bar{\alpha}\rvert^2$, is
small. Thus, the variation of $ I ( z_0,t; q+L)$ can be approximated
by
\begin{align}\label{dz}
 \Delta  I ( z_0,t; q+L)&=\Delta I ( z_0,t; L) +\langle\,
\partial_L I (z_0,t;L) \,\rangle_{\bar{\alpha}}\,\Delta q (t_{\rm ret}) +   \langle \,q(t_{\rm ret})\,\rangle_{\bar{\alpha}}\,\Delta \partial_L I(z_0,t;L)\notag\\
    &\qquad+\langle\, q (t_{\rm ret}) \,\rangle_{\bar{\alpha}}\,\langle\,
\partial^2_L I(z_0,t;L)\,\rangle_{\bar{\alpha}}\,\Delta q(t_{\rm ret})+\frac{1}{2}\,\langle \,q(t_{\rm ret})\,\rangle^2_{\bar{\alpha}}\,\Delta
\partial^2_L I(z_0,t;L)\notag\\
&\qquad\qquad\qquad\qquad+ \mathcal{O}(\Delta^2)\,,
\end{align}
if, given an operator $O$, its variation $\Delta O$ is defined as
$\Delta O=O-\langle O\rangle_{\bar{\alpha}}$. This gives a
leading-order effect of the noise. Later we will offer an
order-of-magnitude estimate of the ignored terms. The first term of
\eqref{dz} arises from the intrinsic fluctuations of the incident
state. For a large value of $\lvert\bar{\alpha}\rvert^2$, the particle number
has relatively small uncertainty; thus according to \eqref{I_exp},
the main contribution to the variation of $ I ( z_0,t; L)$ for fixed
$\bar{\omega}$ does not come from amplitude uncertainty, but is
given by the phase fluctuations, as will be seen later in
Eq.~\eqref{E:kankjcnka}. It is thus associated with shot noise. The
other terms may be understood as the induced effects due to either
the position fluctuations of the mirror, or the modified field
fluctuations as a result of the mean displacement of the mirror.
Then the variation of $I(z_0,t;q+L)$ allows us to define the overall
uncertainty of the effective distance between the detector and the
mirror,
\begin{equation}\label{E:deltaz}
    \Delta z = \frac{\Delta I ( z_0,t; q+L)}{\lvert\langle\,\partial_L I(z_0,t;L) \,\rangle_{\bar{\alpha}}\rvert}\,
    ,
\end{equation}
as long as $ \lvert \, \bar{\omega} \Delta z \,
\rvert < 1$. The normalization factor $\langle \,
\partial_L I(z_0,t;L) \,\rangle_{\bar{\alpha}}$ is to measure the change of $I(z_0,t)$ due to variation of the mirror's position. Control of the above uncertainty is essential to improve the sensitivity of the interferometer, in which the measurement of the separation between the mirrors and the beam splitter plays an important role.

To compute the square of the effective displacement uncertainty $(\Delta z)^{2}$ of the mirror, we may encounter terms such as $ \langle\,S(x)\,S(x')\,\rangle-\langle\,S (x)\,\rangle\,\langle\,S (x')\,\rangle $, $\langle\,T_{zz}(x)\,T_{zz}(x')\,\rangle - \langle \,T_{zz} (x)\,\rangle\,\langle\,T_{zz}(x')\,\rangle $, and their cross correlation terms. Since both $S(x)$ and $T_{zz} (x) $ are quadratic in field operators $\phi(x)$, they are not well-defined in the coincidence limit $x\rightarrow x'$. We expect that the square of the effective displacement uncertainty should contain terms involving a product of four field operators. Thus it may be more transparent if we decompose such a product as follows,
\begin{align}\label{E:wicks}
\phi_1 \phi_2 \phi_3 \phi_4 &= : \phi_1 \phi_2 \phi_3 \phi_4 : + :
\phi_1 \phi_2 : \langle \,\phi_3 \phi_4 \,\rangle_0  \,+ : \phi_1
\phi_3:
\langle\, \phi_2 \phi_4 \,\rangle_0 \notag\\
&\qquad+ :\phi_1 \phi_4: \langle\, \phi_2 \phi_3 \,\rangle_0 \, +
:\phi_2 \phi_3:  \langle \, \phi_1 \phi_4 \, \rangle_0 \, + :\phi_2
\phi_4: \langle \,
\phi_1 \phi_3 \, \rangle_0  \notag\\
&\qquad\qquad+ :\phi_3 \phi_4 : \langle \, \phi_1 \phi_2 \,
\rangle_0 \, + \langle\, \phi_1 \phi_2 \, \rangle_0 \langle\, \phi_3
\phi_4 \, \rangle_0 \, + \langle \,
\phi_1 \phi_3 \, \rangle_0 \langle\, \phi_2 \phi_4 \,\rangle_0 \notag\\
&\qquad\qquad\qquad+ \langle \, \phi_1 \phi_4 \,\rangle_0
\,\langle\, \phi_2 \phi_3 \,\rangle_0 \, ,
\end{align}
where $ \langle\cdots\rangle_0$ denotes the expectation value of the
scalar field in the Minkowski
vacuum. The first term is a fully normal-ordered term, the next six
terms are cross terms, and the final three terms are pure vacuum
terms. For
a single-mode coherent state, the fully normal-ordered terms cancel,
and then we are left with
\begin{align*}
\langle \, \phi_1 \phi_2 \phi_3 \phi_4 \,\rangle_{\bar{\alpha}}
-\langle \,\phi_1 \phi_2  \,\rangle_{\bar{\alpha}} \, \langle\, \phi_3
\phi_4 \, \rangle_{\bar{\alpha}}  &= \langle\, :\phi_1 \phi_3 :
\,\rangle_{\bar{\bar{\alpha}}} \, \langle \, \phi_2 \phi_4
\, \rangle_0 + \text{other cross terms} \\
&\qquad+\langle \,\phi_1 \phi_2 \, \rangle_0 \, \langle \, \phi_3
\phi_4 \, \rangle_0 +\text{other pure vacuum terms} \, .
\end{align*}
The cross terms and the pure vacuum terms may be singular because
the field operators will be evaluated at the same point. It has been
shown that the spacetime average of the pure vacuum term leads to a
result that varies as an inverse power of the size of the
spacetime~\cite{WUF}. On  contrary, the cross
terms depend on the particle density of the coherent state. For a
large value of the particle density, the cross terms can dominate
over the pure vacuum terms~\cite{WUF}. Thus we will consider the
contributions from the cross terms and ignore the pure vacuum terms.
Since the integrals that contain cross terms may still diverge, a
finite result is obtained~\cite{WUF} by choosing the principal value
of the singular integral. Here we take an alternative approach to
cope with the singular vacuum contribution so that all results
become regular. They prove to be the same in the
end.

It is straightforward to compute the expectation value of the
operators in Eq.~\eqref{dz}. Let us focus on  the
large $t \gg L-z_0 $ limit. They are given by
\begin{equation}
\langle \, q(t) \, \rangle_{\bar{\alpha}} =
\lvert\bar{\alpha}\rvert^2
\frac{\mathcal{A}_{\parallel}}{(2\pi)^{3}}\frac{\bar{\omega}}{m}
\times t^2 \, .   \label{q}
\end{equation}
The results of $ \langle \,\partial_L  I \, \rangle_{\bar{\alpha}}$
and $ \langle \,\partial^2_L  I\, \rangle_{\bar{\alpha}}$ can be
obtained by taking the derivative of the
expression~Eq.~\eqref{I_exp} with respect to $L$.

Let the dispersion associated with $\Delta z$ be defined as
\begin{equation}
    \langle\Delta z^{2}\rangle=\langle(\Delta z)^{2}\rangle=\frac{\langle\Delta I^{2}( z_0,t; q+L)\rangle_{\bar{\alpha}}}{\langle\,\partial_L I_T (z_0,t;L) \,\rangle_{\bar{\alpha}}^{2}}\,.
\end{equation}
From \eqref{dz}, the whole expressions in $\langle\Delta
z^{2}\rangle$ can be grouped according to their physical meaning.
They respectively come from the shot noise (sn) contribution associated with intrinsic fluctuations of the incident
coherent fields, and the contributios of radiation pressure
fluctuations (rp) and modified field fluctuations (mf), both of which are
induced by the mirror's motion,
\begin{align}
    \langle\Delta z^{2}\rangle_{sn}&=\frac{\langle\,\Delta I^{2}(z_0,t; L)\,\rangle_{\bar{\alpha}}}{\langle\,\partial_L I (z_0,t;L) \,\rangle_{\bar{\alpha}}^{2}}\,,\label{E:sn}\\
    \langle\Delta z^{2}\rangle_{rp}&=\langle\Delta q^{2}(t)\rangle_{\bar{\alpha}}\,,\label{E:rr}\\
    \langle\Delta z^{2}\rangle_{mf}&=\langle\,q(t)\,\rangle^{2}_{\bar{\alpha}}\; \frac{\langle\,\Delta(\partial_{L}I)^{2}( z_0,t; L)\,\rangle_{\bar{\alpha}}}{\langle\,\partial_L I(z_0,t;L)
    \,\rangle_{\bar{\alpha}}^{2}}\,.\label{E:mf}
\end{align}
In addition and more importantly, there exist the cross terms owing
to correlation between different sources of
uncertainty. Up to the order $q^2$, we lump these terms
together in
\begin{align}\label{z_cor}
    \langle\Delta z^{2}\rangle_{cor}&=\frac{1}{\langle\,\partial_L I(t)\, \rangle_{\bar{\alpha}}} \, \langle\,\bigl\{\Delta I(t)  \,, \Delta q(t) \bigr\}\,\rangle_{\bar{\alpha}}+\frac{ \langle\,q(t)\,\rangle_{\bar{\alpha}}}{\langle\,\partial_L I(t)\,\rangle^2_{\bar{\alpha}}} \,\langle\,\bigl\{\,\Delta I(t)\,, \Delta \partial_L I(t)\bigr\}\,\rangle_{\bar{\alpha}}\nonumber \\
    &\qquad+ \frac{\langle\,q(t)\,\rangle_{\bar{\alpha}}}{\langle\,\partial_LI(t)\,\rangle_{\bar{\alpha}}}\,\langle\,\bigl\{\Delta\partial_LI(t)\,,\Delta q(t)\bigr\}\,\rangle_{\bar{\alpha}}+ \frac{\langle\,q(t)\,\rangle_{\bar{\alpha}}\langle\,\partial^2_LI(t)\,\rangle_{\bar{\alpha}} }{\langle\,\partial_LI(t)\,\rangle^2_{\bar{\alpha}}} \,\langle\,\bigl\{\Delta I(t) \,,\Delta q(t)\bigr\}\,\rangle_{\bar{\alpha}}\notag\\
    &\qquad\qquad\qquad+\frac{1}{2}\frac{\langle\,q(t)\rangle^2_{\bar{\alpha}}}{\langle \,\partial_LI(t)\,\rangle^2_{\bar{\alpha}}}\,\langle\,\bigl\{\Delta\partial^{2}_LI(t)\,,\Delta I(t)\bigr\}\,\rangle_{\bar{\alpha}} \,.
\end{align}
From now afterwards, the quantity $I$ and its derivatives are
understood to be evaluated at $z=L$, and we replace $I(z_{0},t;L)$
simply by its shorthand notation $I(t)$. We will discuss these
contributions individually in the following sections. In particular,
the emphasis will be put on the correlation effects that may give
the hope to reduce the overall uncertainty. Here we note that
Eqs~\eqref{E:sn}, \eqref{E:rr}, \eqref{E:mf}, \eqref{z_cor} involve
the expectation values of the anticommutators of the operators, so
in the leading-order approximation we will obtain the same results
irrespective of operator ordering on expanding Eq.~\eqref{I_L}.

\subsection{shot-noise terms}
To the overall dispersion of $\Delta z$, the contribution from the
intrinsic fluctuations of the incident field is given by the term
$\langle\,\Delta I^2\,\rangle_{\bar{\alpha}} $, which is expressed as
\begin{align}
    &\quad\langle \, \Delta I^2(z_0,t;L) \, \rangle_{\bar{\alpha}}= \langle\,I^2(z_0,t;L)\,\rangle_{\bar{\alpha}} -\langle\, I(z_0,t;L)\,\rangle^2_{\bar{\alpha}}
\notag\\
    &=\int_0^t\!ds\!\int\! d\mathcal{A}_{\parallel}\!\int'\!\frac{d^3 k}{( 2
\pi)^{\frac{3}{2}}}\!\int'\!\frac{d^3 k'}{( 2 \pi)^{\frac{3}{2}}}\!\int_0^t\!d s'\!\int d\mathcal{A}'_{\parallel}\!\int'\!\frac{d^3\tilde{k}}{( 2 \pi)^{\frac{3}{2}}}\!\int'\!\frac{d^3 \tilde{k}'}{( 2\pi)^{\frac{3}{2}}}\;\frac{1}{\sqrt{2 k}} \, \frac{1}{\sqrt{2 k'}}\;
\frac{1}{\sqrt{2 \tilde{k}}}\, \frac{1}{\sqrt{2 \tilde{k}'}}\notag\\
    &\qquad\qquad\qquad\qquad \times
{\cal{U}}^*_{\mathbf{k}} ( \mathbf{x}_{\parallel}, z_0 ,
 s;L)  \,{\cal{U}}_{\mathbf{k'}} ( \mathbf{x}_{\parallel}, z_0,
s;L)\, {\cal{U}}^*_{\tilde{\mathbf{k}}} ( \mathbf{x}'_{\parallel}, z_0 ,
, s';L)  \, {\cal{U}}_{\tilde{\mathbf{k}}'} ( \mathbf{x}'_{\parallel}, z_0,
s';L) \notag\\
    &\qquad\qquad\qquad\qquad\qquad\qquad\qquad\times\bigg[ \langle a^{\dagger}_{\mathbf{k}}  a_{ \mathbf{k}'}^{\vphantom{\dagger}}
a^{\dagger}_{ \tilde{\mathbf{k}}} a_{ \tilde{\mathbf{k}}'}^{\vphantom{\dagger}} \rangle_{\bar{\alpha}} - \langle a^{\dagger}_{\mathbf{k}}
a_{ \mathbf{k}'}^{\vphantom{\dagger}}\rangle_{\bar{\alpha}}  \langle a^{\dagger}_{\tilde{\mathbf{k}}} a_{\tilde{\mathbf{k}}'}^{\vphantom{\dagger}}\rangle_{\bar{\alpha}}
 \bigg]
\notag \\
 &= \frac{\mathcal{A}_{\parallel}}{(2\pi)^{3}}\, \lvert\bar{\alpha}\rvert^2\frac{8}{ \bar{\omega}}\,\sin^2[\bar{\omega}(L-z_0)] \int_0^t ds
\int_0^t ds' \int_0^{\infty} \frac{d\omega}{2\pi} \frac{1}{2\omega}
\sin^2[\omega(L-z_0)] e^{-i ( \omega-\bar{\omega}) s} e^{ i( \omega-\bar{\omega}) s'} \, ,\notag
\end{align}
where the area integral is carried out over a measuring plane at $z=z_0$, and
its total area $\mathcal{A}_{\parallel}$ is assume to be large but
finite. Performing the time integrations and using the fact that
\begin{equation}
\lim_{ t \rightarrow \infty} \frac{1}{\omega-\bar{\omega}} \sin \left[ \frac{(\omega-\bar{\omega})  t}{2}\right] =\pi
\delta (\omega-\omega_0) \,, \label{larget}
\end{equation}
we have
\begin{equation}
\langle \, \Delta I^2 (z_0,t;L) \, \rangle_{\bar{\alpha}}\approx
\frac{\mathcal{A}_{\parallel}}{(2\pi)^{3}}\, \lvert\bar{\alpha}\rvert^2\frac{4}{ \bar{\omega}^{2}}\,\sin^4[\bar{\omega}(L-z_{0})] \times t \, .
\end{equation}
Together with the normalization factor $\langle \,
\partial_L I \, \rangle_{\bar{\alpha}}^2$, Eq.~\eqref{E:sn} gives
the positive contribution
\begin{equation}
    \langle\Delta z^{2}\rangle_{sn}=\frac{\langle \, \Delta I^2 \, \rangle_{\bar{\alpha}}}{\langle \,
\partial_L I \, \rangle_{\bar{\alpha}}^2} \approx  \frac{1}{P \bar{\omega}
t} \frac{1}{4} \tan^2[\bar{\omega}(L-z_0)] \,  \,,\quad\quad t \gg
1/\bar{\omega}\,,\quad\text{and}\quad t\gg L-z_{0}\,, \label{sn2}
\end{equation}
where the energy flux $P$ of the incident state is defined by
\begin{equation}
    P=\frac{\mathcal{A}_{\parallel}}{(2\pi)^{3}}\lvert\bar{\alpha}\rvert^2\bar{\omega}\,.
    \label{P}
\end{equation}
The result \eqref{sn2} is inversely proportional to the power $P$,
and that is the typical behavior of the shot nose.

Since the read-out field $I (z_0,t; L)$ is to count
the number of the particles being detected, the
same result can be obtained in terms of the particle number operator
$n$ if we make such an identification,
\begin{equation}
    I (z_0,t; L) =  \frac{ 2}{\bar{\omega}} \sin^2 [\bar{\omega} ( L
    -z_0) \label{I_n}
]\times  n \,,
\end{equation}
and notice that for a coherent state, $\langle\,n^2\,\rangle_{\bar{\alpha}}-\langle\,n\,\rangle_{\bar{\alpha}}^2=\langle\,n\,\rangle_{\bar{\alpha}}$, where the mean number of particles $\langle\,n\,\rangle_{\bar{\alpha}}$ that reaches the surface
$\mathcal{A}_{\parallel}$ within time $t$ is that $\langle\,n\,\rangle_{\bar{\alpha}}=\frac{\mathcal{A}_{\parallel}}{(2\pi)^{3}}\,\lvert\bar{\alpha}\rvert^{2}\,t$. This shot noise can be understood as phase fluctuations of the field~\cite{SCU}. The phase uncertainty of $I$ can be effectively
defined as
\begin{equation}\label{E:kankjcnka}
    \Delta \phi=  \frac{\sqrt{ \langle\, \Delta^2 I (\phi,t)\, \rangle_{\bar{\alpha}}}}{ \lvert \langle \partial_{\phi} I (\phi,t) \rangle_{\bar{\alpha}} \rvert } \,.
\end{equation}
The variance of the phase $ \Delta \phi $ is found proportional to
$1/P$ as well. It is worth to mention that this result depends on
not only the input power $P$ but also the distance between the
mirror and detector due to interference between the incident and
reflected fields. The proper choice of the parameter $L-z_0$ can be
used to avoid the potentially large uncertainty due to the factor
$\tan^2[\bar{\omega}(L-z_0)]$ and then to minimize the overall
uncertainty. The same strategy has been used by
Caves in~\cite{CA1,CA2} to finding the appropriate spot in the
fringe pattern so as to further reduce the overall uncertainty in
the interferometer.

\subsection{radiation-pressure-fluctuations terms}
Next, we consider the contribution to $\langle\,\Delta z^{2}\,\rangle$ from the radiation pressure fluctuations. This contribution is given by
$\langle \, \Delta q^{2}(t)\,\rangle_{\bar{\alpha}}$:
\begin{equation}
    \langle\,\Delta z^{2}\,\rangle_{rp}=\langle \, \Delta q^{2}(t)\,\rangle_{\bar{\alpha}}=
\langle\, q^2 \,\rangle_{\bar{\alpha}}-\langle\, q
\,\rangle^2_{\bar{\alpha}} \,,
\end{equation}
where the most dominant contribution is found to be
\begin{align}\label{rpfa+a}
\langle \, \Delta q^2 \,\rangle_{\bar{\alpha}} &= \int_0^{t}\! d\tau
\!\int_0^{\tau}\! ds\! \int d\mathcal{A}_{\parallel} \int_0^{t}\!
d\tau'\!\int_0^{\tau'}\!
ds'\! \int\!d\mathcal{A}'_{\parallel} \notag \\
&\qquad\qquad\int'\!\frac{d^3 k }{( 2 \pi)^{\frac{3}{2}}}\!\int'\!
\frac{d^3 k' }{( 2 \pi)^{\frac{3}{2}} }\! \int'\! \frac{d^3\tilde{k}}{( 2 \pi)^{\frac{3}{2}}}\!\int'\!\frac{d^3 \tilde{k}'}{( 2
\pi)^{\frac{3}{2}} } \!\frac{1}{\sqrt{2 k}}\frac{1}{\sqrt{2 k'}}
\,\frac{1}{\sqrt{2 \tilde{k}}}\frac{1}{\sqrt{2 \tilde{k}'}}
\notag\\
&\biggl\{  \partial_{z_0 } {\cal{U}}^*_{\mathbf{k}} (
\mathbf{x}_{\parallel}, z_0 , s)
\partial_{z_0}\, {\cal{U}}_{\mathbf{k}'} ( \mathbf{x}_{\parallel}, z_0, s) \, \partial_{z_0 } {\cal{U}}^*_{\tilde{\mathbf{k}}} ( \tilde{\mathbf{x}}_{\parallel}, z_0 , s')  \partial_{z_0}
{\cal{U}}_{\tilde{\mathbf{k}}'} (
\tilde{\mathbf{x}}_{\parallel}, z_0, s') \biggr.\notag \\
&\qquad\qquad\qquad\qquad\times \bigg[ \langle a^{\dagger}_{k} a_{
k'} a^{\dagger}_{\tilde{k}} a_{ \tilde{k}'} \rangle_{\bar{\alpha}} -
\langle a^{\dagger}_{k} a_{
k'}\rangle_{\bar{\alpha}} \langle a^{\dagger}_{\tilde{k}} a_{\tilde{k}'}\rangle_{\bar{\alpha}} \bigg]\notag \\
&+\partial_{z_0 } {\cal{U}}^*_{\mathbf{k}'} (
\mathbf{x}_{\parallel}, z_0 , s)
\partial_{z_0}\, {\cal{U}}_{\mathbf{k}} ( \mathbf{x}_{\parallel},z_{0},s) \, \partial_{z_0 } {\cal{U}}^*_{\tilde{\mathbf{k}}'} ( \tilde{\mathbf{x}}_{\parallel}, z_{0} , s')  \partial_{z_0}
{\cal{U}}_{\tilde{\mathbf{k}}} (
\tilde{\mathbf{x}}_{\parallel},z_{0}, s') \notag \\
&\qquad\qquad\qquad\biggl.\times\,\bigg[ \langle
a^{\dagger}_{k'} a_{ k} a^{\dagger}_{\tilde{k}'} a_{ \tilde{k}}
\rangle_{\bar{\alpha}} - \langle a^{\dagger}_{k'} a_{
k}\rangle_{\bar{\alpha}} \langle a^{\dagger}_{\tilde{k}'}
a_{\tilde{k}}\rangle_{\bar{\alpha}} \bigg] + \text{c.c.}
\biggr\}_{z_{0}=L}+\text{subdominant
terms}\notag  \\
&= \frac{4\bar{\omega}}{m^2}
\frac{\mathcal{A}_{\parallel}}{(2\pi)^{3}}\,\lvert\bar{\alpha}\rvert^2\int_0^t\!d\tau\!\int_0^{\tau}\!ds\!\int_0^t\!d{\tau'}\!\int_0^{\tau'}\!ds'\!\int_0^{\infty}\!\frac{d\omega}{2\pi}
\;\omega\,e^{-i (\omega-\bar{\omega}) s} \, e^{
i(\omega-\bar{\omega}) s'}\notag\\
&\qquad\qquad\qquad\qquad\qquad\qquad\qquad+\text{subdominant
terms} \,\notag\\
&\approx\frac{\lvert\bar{\alpha}\rvert^2\bar{\omega}}{m^2}
\frac{\mathcal{A}_{\parallel}t^{2}}{(2\pi)^{3}}\,\bar{\omega}\,t=  \frac{P \bar{\omega} t^3}{m^2}\,.
\end{align}
This result is valid for large time. All other subdominant contributions will be discussed in Appendix~\ref{S:deltaq}.
The result of $ \langle \, \Delta q^2
\,\rangle_{\bar{\alpha}}$ at large time can be equivalently obtained
by the particle-number operator approach, and it is given by the time
integration of the velocity operator of the form
\begin{equation}
v = \frac{2\bar{\omega}}{m}\,n \, , \label{v_n}
\end{equation}
where $n$ is the particle number operator.  In other words,
the momentum received by the
mirror can be effectively given by the momentum
transfer of the scalar particles when they elastically bounce off the mirror.

In contrast to the shot-noise type contribution that has the
$P^{-1}$ dependence, the stochastic motion of the mirror due to
radiation pressure fluctuations yields an uncertainty proportional
to the incident power $P$. This uncertainty can be
attributed to the amplitude fluctuations of the incident field.
These two types of noise have been extensively discussed
in~\cite{CA1,CA2,LO} in the context of the interferometer. In
particular, it is stated in~\cite{JA} that shot noise given by the
photon counting error can be understood in terms of phase
fluctuations of the incident light while radiation pressure
fluctuations are related to amplitude fluctuations. Our results are
consistent with the arguments in~\cite{CA1,CA2,LO,JA}.

Other than contributions related to well-known shot noise and
radiation-pressure fluctuations, modified field fluctuations also
brings in a new component of noise. As the radiation pressure pushes
the mirror into motion, the scalar field in the proximity of the
mirror must change accordingly at every moment to enforce the
boundary condition. In turn, this modification of field fluctuations
will manifest themselves as quantum noise. Since these fluctuations
arise naturally from the moving boundary condition of the fields, it
must be included consistently in the expansion of Eq.~\eqref{dz},
and this adds a new contribution into $\Delta z^{2}$, as is
described by \eqref{E:mf}. With the help of Eq.~\eqref{q} and
\begin{equation*}
    \langle\,\Delta(\partial_LI)^{2} (z_0,t;L)\,\rangle_{\bar{\alpha}}=\langle\,(\partial_L I)^{2}\,\rangle_{\bar{\alpha}}-\langle\,\partial_LI\,\rangle^2_{\bar{\alpha}}\approx 4\,\frac{\mathcal{A}_{\parallel}}{(2\pi)^{3}}\,\lvert\bar{\alpha}\rvert^2\sin^2[2\bar{\omega}(L-z_0)]\times t\,,
\end{equation*}
we have
\begin{equation}
    \langle\,\Delta z^{2}\,\rangle_{mf}=\frac{ \langle\, q\, \rangle^2_{\bar{\alpha}}}{\langle\,\partial_L I\,\rangle^2_{\bar{\alpha}}} \langle\,\Delta^2 \partial_L I\,\rangle_{\bar{\alpha}} \approx \frac{P\bar{\omega} t^3}{m^2}\,,
\end{equation}
for $ \bar{\omega}t \gg 1$. Apparently, this term has mixed features
that involve $\langle q \rangle_{\alpha} $ and
$\langle\,\partial_{L}I\,\rangle_{\bar{\alpha}}$. Hence it may
depend on both the incident power and the phase of incident
state. In the end, however, the phase dependence
cancels in this configuration so that the noise from modified field
fluctuations depends only on the power of the input state.

\subsection{correlation terms}
In this framework, it is quite straightforward to
compute the correlation between various noise. The correlation
between shot noise and noise from random motion is given by
\begin{equation*}
    \frac{1}{\langle\,\partial_L I\, \rangle_{\bar{\alpha}}} \, \langle\,\bigl\{\Delta I  \,, \Delta q \bigr\}\,\rangle_{\bar{\alpha}}\approx
    \frac{t}{m}\,\tan[\bar{\omega}(L-z_0)]\,,
\end{equation*}
which can take either sign, depending on
the distance between the detector and the mirror's original
position. The presence of such sign-vary expressions gives the hope
to reduce the overall uncertainty $\langle\,\Delta
z^{2}\,\rangle_{\bar{\alpha}}$. It is worth to
mention that the same result can be
obtained by the particle-number operator approach using
Eqs.~\eqref{I_n}, and~\eqref{v_n}.

There exists another correlation term due to the
additional source of noise, that is, the correlation between shot
noise and the noise due to modified field fluctuation,
\begin{equation*}
    \frac{ \langle\,q\,\rangle_{\bar{\alpha}}}{\langle\,\partial_L I\,\rangle^2_{\bar{\alpha}}} \,\langle\,\bigl\{\,\Delta I\,, \Delta \partial_L I\bigr\}\,\rangle_{\bar{\alpha}}\approx \frac{t}{m} \, \tan[\bar{\omega}(L-z_0)]\,,
\end{equation*}
which has the same order of the magnitude as the
the previous result. Furthermore, it leads to
\begin{align*}
    &\quad\frac{\langle\,q\,\rangle_{\bar{\alpha}}}{\langle\,\partial_L I\,\rangle_{\bar{\alpha}}} \, \langle\,\bigl\{\Delta \partial_LI\,, \Delta q \bigr\}\,\rangle_{\bar{\alpha}}+\frac{\langle\,q\,\rangle_{\bar{\alpha}}\langle\,\partial^2_L I\,\rangle_{\bar{\alpha}}}{\langle\,\partial_L I\,\rangle^2_{\bar{\alpha}}} \,\langle\,\bigl\{\Delta I \,, \Delta q\bigr\}\,\rangle_{\bar{\alpha}}+\frac{1}{2} \frac{\langle\,q\rangle^2_{\bar{\alpha}}}{\langle\,\partial_L I\,\rangle^2_{\bar{\alpha}}} \,\langle\,\bigl\{\Delta \partial^{2}_L
I\,, \Delta I\bigr\}\,\rangle_{\bar{\alpha}}\notag\\
    &\approx \frac{P \bar{\omega} t^3}{m^2}\left(\frac{7}{2}-\frac{3}{2} \tan^2 [\bar{\omega}(L-z_0)]
    \right)\,,\qquad\qquad\qquad \bar{\omega}t\gg 1\,,
\end{align*}
when we add up correlation terms of the order
$q^2$, as is summarized within the expression
$\langle\,\Delta^{2}z\rangle_{cor}$ in Eq. \eqref{z_cor}.

Now we may put all effects together, and the
overall uncertainty $\langle \, \Delta z^2 \,
\rangle_{\bar{\alpha}}$ is then given by
\begin{equation}
    \langle \, \Delta z^2 \, \rangle_{\bar{\alpha}} =
\frac{1}{P\bar{\omega} t} \frac{\zeta^2}{4} + \frac{t}{m} 2 \zeta
+\frac{P \bar{\omega} t^3}{m^2} \left( \frac{11}{2} -\frac{3}{2}
\zeta^2 \right) \, , \label{dz2}
\end{equation}
where $\zeta=\tan[\bar{\omega}(L-z_{0})]$. Interference between the
incoming and reflected waves gives the quantity $I$
the dependence on the effective distance between the mirror and
detector. Later, the correlation between
different sources of quantum noise can be found to possibly reduced the overall quantum noise.




\section{effect of correlation}\label{sec3}
In the case of single-mode coherent state, $\zeta>0$ represents
positive correlation between shot noise and radiation-pressure
fluctuations, while $\zeta<0$ denotes negative correlation. The
minimum value of $\langle\,\Delta z^{2}\,\rangle_{\bar{\alpha}}$ is
found to be
\begin{equation}\label{E:fnhw1}
    \min\,\langle\,\Delta z^{2}\,\rangle_{\bar{\alpha}}=\zeta\left( 2\pm \sqrt{\frac{11}{2} -\frac{3}{2}\,\zeta^2} \right) \frac{t}{m}\,,
\end{equation}
with an optimal value of the power $P$ given by
\begin{equation}\label{E:fnhw2}
    P_{\rm opt}=\pm\frac{\zeta}{\sqrt{ 22-6 \zeta^2}} \frac{m}{\bar{\omega}t^2}\,.
\end{equation}
In Eqs.~\eqref{E:fnhw1} and \eqref{E:fnhw2}, the ``+ (-)'' sign
corresponds to the positive (negative) $\zeta$ case.

\begin{figure}
\centering
    \scalebox{0.8}{\includegraphics{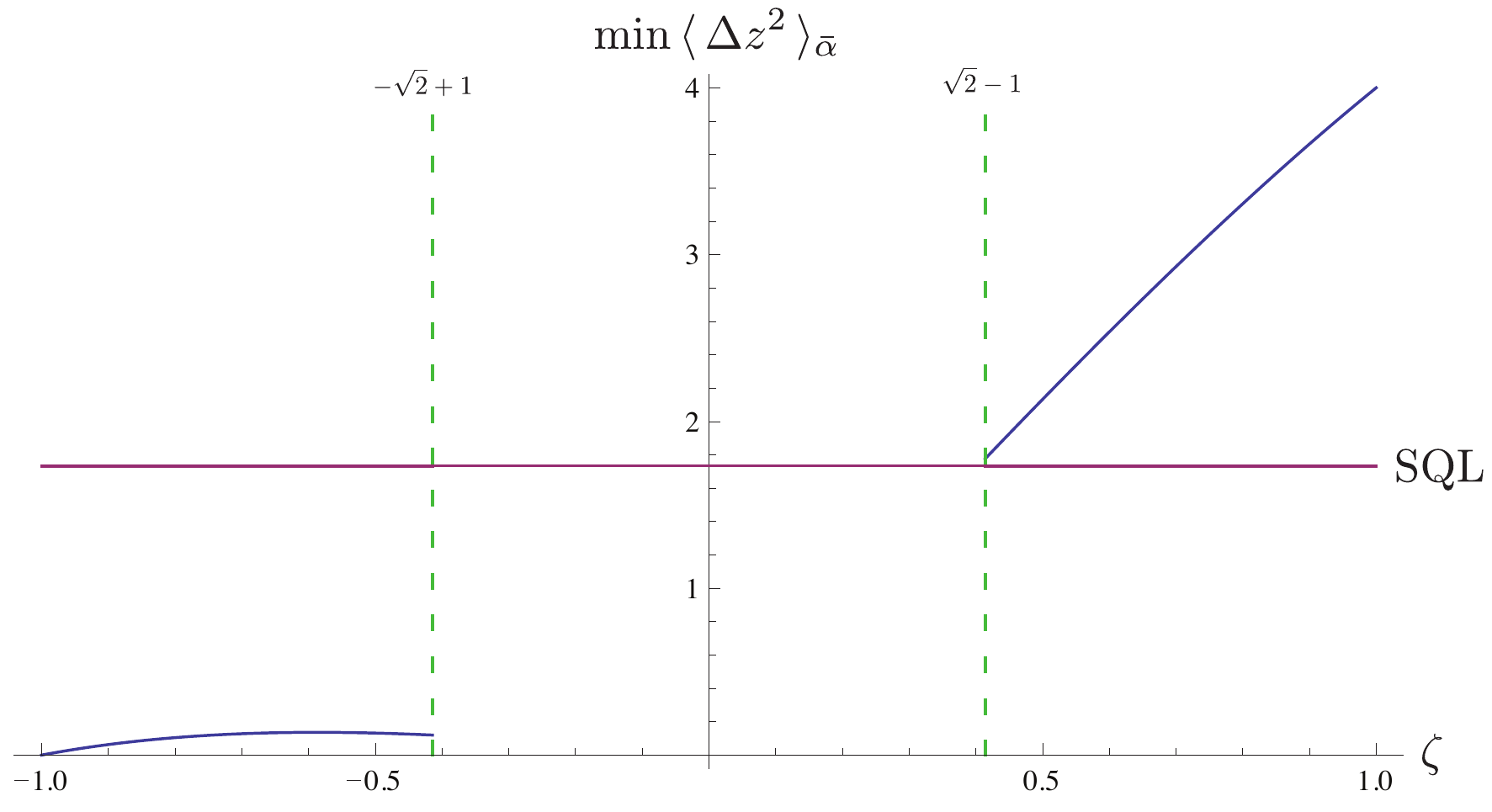}}
    \caption{The  minimal value of
$\langle\,\Delta z^{2}\,\rangle_{\bar{\alpha}}$ as a function of
$\zeta$ in comparison with the SQL. The $\min\,\langle\,\Delta
z^{2}\,\rangle_{\bar{\alpha}}$ is drawn in the unit of
$\frac{t}{m}$. }\label{Fi:deltaz_xi}
\end{figure}

The minimal value of $\langle\,\Delta
z^{2}\,\rangle_{\bar{\alpha}}$ depends on $\zeta$, but the range of
the values of $\zeta$ can not be arbitrary; instead, it has to be
determined in a consistent way with the underlying assumptions. Slow
motion of the mirror, ${\bar{\omega}} q\ll1$, can be translated into
an inequality, $P_{\rm opt}\,\bar{\omega} t^2/m\ll1$, as can be seen
from \eqref{q} and \eqref{P}. The latter in turn implies $\left|
\zeta \right| < 1$ by substituting the expression of $P_{\rm opt}$
in \eqref{E:fnhw2} directly into the inequality.  On the other hand,
the definition of the effective displacement of the mirror in
\eqref{E:deltaz}, formally by $\Delta I\approx\Delta
z\times\partial_{L}I$, implicitly assumes
\begin{equation*}
    \lvert\,\Delta z\times\partial_{L}I\,\rvert > \lvert\,\frac{1}{2}\,(\Delta z)^{2}\partial^{2}_{L}I\,\rvert\,.
\end{equation*}
Together with the slow motion assumption, it gives
\begin{equation*}
   \lvert\,\bar{\omega}\Delta z\,\rvert < \frac{2 \left| \zeta \right| }{1-\zeta^{2}}< 1\,.
\end{equation*}
The last inequality is imposed to ensure that the range of the value
of $\Delta z$ is consistent with nonrelativistic motion of the
mirror as well as the requirement on the definition of $\Delta z$ in~\eqref{E:deltaz}, leading to
$\left|\zeta\right|>\sqrt{2}-1\sim0.414$. Thus $\left|\zeta\right|$
should lie within the range
\begin{equation*}
    \sqrt{2}-1 < \left| \zeta \right|<1 \, .
\end{equation*}
The minimal value of $\langle\,\Delta
z^{2}\,\rangle_{\bar{\alpha}}$ as a function of $\zeta$ is shown in
Fig.~\ref{Fi:deltaz_xi}.

For a positive $\zeta$, $\min\,\langle\,\Delta
z^{2}\,\rangle_{\bar{\alpha}}$ is a monotonic function of $\zeta$,
so it varies between
\begin{equation}\label{E:wiownkx}
    1.78\,\frac{t}{m} < \min\,\langle\,\Delta z^{2}\,\rangle_{\bar{\alpha}} < 4\,\frac{t}{m}\,,
\end{equation}
compared with the SQL, $\Delta z^{2}_{\rm SQL}=t/m$.
It is seen that we cannot beat SQL in the positive correlation case. For the upper (lower) bound in \eqref{E:wiownkx},
the corresponding optimal power is given by
\begin{equation*}
    P_{\rm opt}\approx0.25\,\frac{m}{\bar{\omega} t^2}\qquad\qquad\text{and}\qquad\qquad P_{\rm opt}\approx0.09\,\frac{m}{\bar{\omega} t^2}\,,
 \end{equation*}
respectively. On the other hand, for negative $\zeta$,
$\min\,\langle\,\Delta^2z\,\rangle_{\bar{\alpha}}$ is not a monotonic with $\zeta$, having a maximum at $\zeta=-0.588$, so in this case the largest value of
$\min\,\langle\,\Delta^2z\,\rangle_{\bar{\alpha}}$ is
\begin{equation*}
    \min\,\langle\,\Delta z^2\,\rangle_{\bar{\alpha}}\approx 0.136\,\frac{t}{m}\,,
\end{equation*}
with the optimal input power
\begin{equation*}
    P_{\rm opt}\approx0.132\,\frac{m}{\bar{\omega} t^2}\,.
\end{equation*}
In the case of negative correlation, even the largest value of
$\min\,\langle\,\Delta^2z\,\rangle_{\bar{\alpha}}$ can be lower than
the standard quantum limit. Therefore, the value of the position
uncertainty can be further reduced by other choices of $\zeta$, and na\"ively
it can be as small as possible. However, there is a
caveat. We have ignored higher-order terms in the series expansion
\eqref{I_T_q2} and in the calculation of $\langle\,\Delta
z^2\,\rangle_{\bar{\alpha}}$ in terms of small $q$. In
addition, we have employed the leading order approximation in the long-time limit
by considering a large number of the particles in the coherent state. Hence $\min\,\langle\,\Delta z^2\,\rangle_{\bar{\alpha}}$
cannot be arbitrarily small in that the subleading results
will sooner or later contribute comparable effects. At any rate, we have shown that the standard quantum limit
can be possibly overcome by establishing correlation between the
intrinsic fluctuations of the field (shot noise) and its
induced fluctuations due to the dynamics of the mirror (radiation
pressure fluctuations and modified field fluctuations).

So far, we have considered  the single-mode
coherent state. In reality, coherent
state is at best prepared with a
finite frequency bandwidth, so the observed response of the detector
should be averaged over its bandwidth. In general,
let the frequency distribution be described by some function of the
form $f(\omega ,\bar{\omega} ;\sigma_0)$, where $\bar{\omega}$ is
the mean frequency of the band and $2\sigma_{0}$ the bandwidth. The
average over the distribution function is given by
\begin{equation*}
    \overline{O}(\bar{\omega},\sigma_0) =\int_{0}^{\infty}d\omega\; f(\omega ,\bar{\omega}
;\sigma_0) \,  O(\omega) \, .
\end{equation*}
In our case, the quantity $O$ is a product of a fast oscillating function of $\omega$ due to the macroscopic scale $L-z_{0}$ and a relatively slowly varying component over the interval $\bar{\omega}-\sigma_{0}<\omega<\bar{\omega}+\sigma_{0}$. Suppose the bandwith is narrow in the sense that $\sigma_{0}\ll\bar{\omega}$ and $\sigma_{0}(L-z_{0})\gg1$. Then the average can be approximated by substituting the variable $\omega$ in the slowly varying component with the mean frequency $\bar{\omega}$ and directly performing average over the fast oscillating part.

\begin{figure}
\centering
    \scalebox{0.64}{\includegraphics{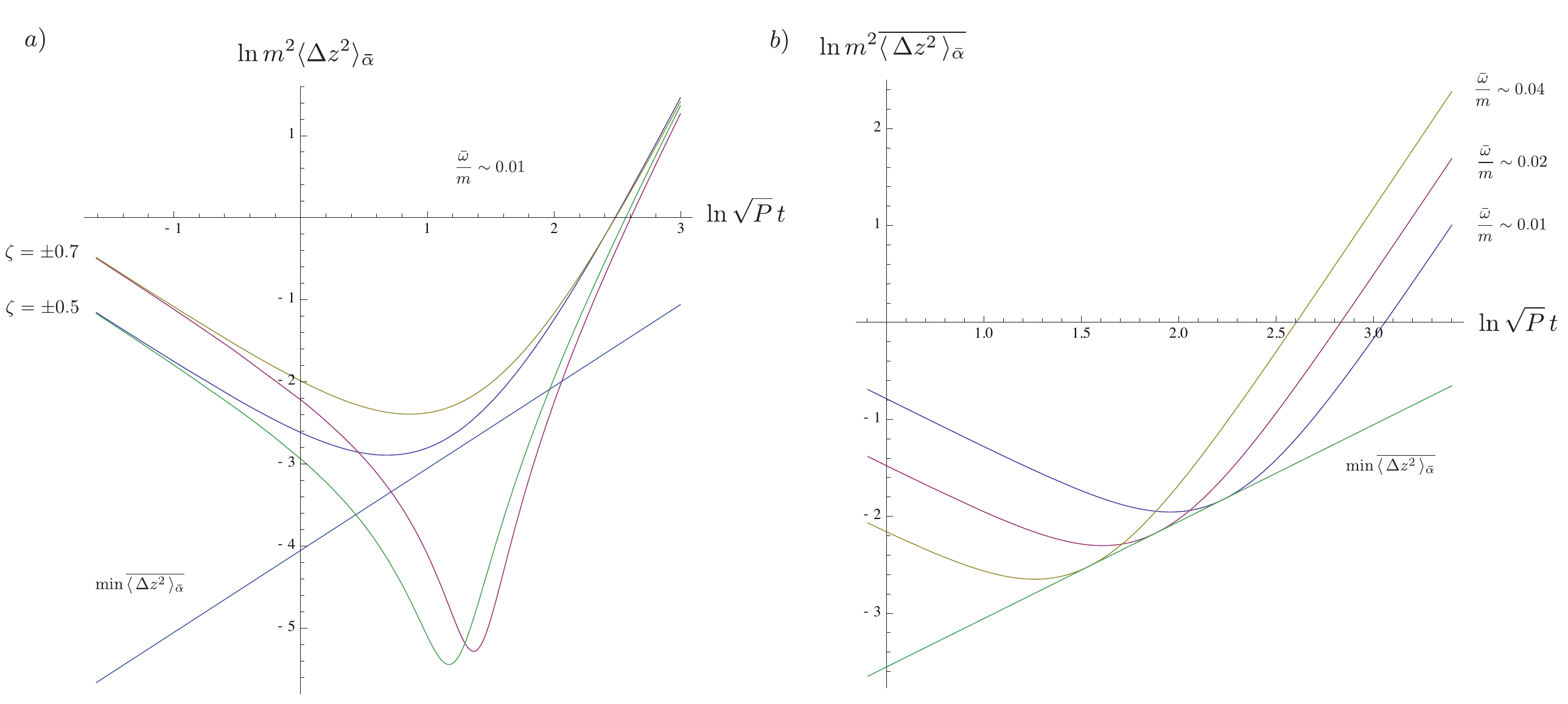}}
    \caption{a) Log-log plot of $\langle\,\Delta
    z^2\,\rangle_{\bar{\alpha}}$ versus $\sqrt{P} t$
     for various values of $\zeta$. The straight line corresponds to the SQL. The parameters
    $\frac{ {\bar{\omega}}}{m} = 10^{-2}$  and $\frac{P}{m^{2}}=10^{-4}$ are chosen. b) Log-log plot of  $\overline{\langle\,\Delta z^{2}\,\rangle_{\bar{\alpha}}}$ versus
     $\sqrt{P} t$ for various values of $\frac{ {\bar{\omega}}}{m} $. The straight line is  the result of $ \min \overline{\langle\,\Delta
     z^{2}\,\rangle_{\bar{\alpha}}}$.
     $\frac{P}{m^{2}}=10^{-4}$ is chosen.}\label{Fi:sql}
\end{figure}

Thus if the coherent parameter $\alpha$ is independent of the frequency, the average of the normalization factor $\langle\,\partial_L I\,\rangle_{\bar{\alpha}}^2 $ is given by
\begin{equation*}
    \overline{\langle\,\partial_LI\,\rangle_{\bar{\alpha}}^2}=4\lvert\bar{\alpha}\rvert^4\,\frac{\mathcal{A}_{\parallel}^2}{(2\pi)^6}\, \overline{\sin^2 [ 2 \omega (L-z_0)]}\times t^2\approx2\lvert \bar{\alpha} \rvert^4\,\frac{\mathcal{A}_{\parallel}^2}{(2\pi)^6}\times t^2\,, \label{sin2_f}
\end{equation*}
from Eq.~\eqref{I_exp}. Likewise, the corresponding frequency averages of the various sources of noise can be computed.
The average of the shot noise term is given by
\begin{equation*}
    \overline{\langle \, \Delta I^2\,
\rangle}_{sn}= 4\lvert\bar{\alpha}\rvert^2\,\frac{\mathcal{A}_{\parallel}}{(2\pi)^3}\frac{1}{\bar{\omega}^{2}} \; \overline{\sin^4 [\omega (L-z_0)] }\times t\approx\lvert\bar{\alpha}\rvert^2\,\frac{\mathcal{A}_{\parallel}}{(2\pi)^3}\,\frac{3}{2\bar{\omega}^2}\times t\,,
\end{equation*}
and the average of the radiation-pressure fluctuations contribution is
\begin{equation*}
    \overline{\langle\,\Delta I^2\,\rangle}_{rp}=\lvert\bar{\alpha}\rvert^6\,\frac{\mathcal{A}^{3}_{\parallel}}{(2\pi)^9}\frac{16\,\bar{\omega}^{2}}{m^{2}}\;\overline{\sin^2 [\omega (L-z_0)]\cos^2 [\omega (L-z_0)] }\times t^{5}\approx\lvert\bar{\alpha}\rvert^6\,\frac{\mathcal{A}^{3}_{\parallel}}{(2\pi)^9}\frac{2\bar{\omega}^{2}}{m^{2}}\times t^{5}\,.
\end{equation*}
It is interesting to note that the average of the rest of the terms totally cancels out by themselves, even
though they contain drastically different information about correlation
between noises and modified field fluctuations,
\begin{equation*}
    \overline{\langle \,\Delta I^2\,\rangle}_{mf}+\overline{\langle\,\Delta I^2\,\rangle}_{cor}=0\,.
\end{equation*}
The presence of finite frequency bandwidth reduces $\langle\,\Delta
I^{2}(z_{0},t;q+L)\,\rangle$ to solely consist of contributions of
shot noise and radiation-pressure fluctuations. Hence
$\overline{\langle\,\Delta z^{2}\,\rangle_{\bar{\alpha}}}$ in the end is given by
\begin{equation}
    \overline{\langle\,\Delta z^{2}\,\rangle_{\bar{\alpha}}}=\frac{\overline{\langle\,\Delta I^{2}\,\rangle_{\bar{\alpha}}}}{\overline{\langle\,\partial_LI\,\rangle_{\bar{\alpha}}^2}}=\frac{3}{4}\frac{1}{P\bar{\omega} t}+\frac{P\bar{\omega} t^3}{m^2} \,,
\end{equation}
which can be compared with Eq.~\eqref{dz2}.
Minimization of $\overline{\langle\,\Delta
z^{2}\,\rangle_{\bar{\alpha}}}$ can be achieved with an optimal
value of the power $P$,
\begin{equation}
    P_{\rm opt}=\frac{\sqrt{3}}{2} \frac{m}{\bar{\omega} t^2}\,,
\end{equation}
leading to the minimal value of $\overline{\langle\,\Delta z^{2}\,\rangle_{\bar{\alpha}}}$,
\begin{equation}
    \min \overline{\langle\,\Delta z^{2}\,\rangle_{\bar{\alpha}}}=\sqrt{3}\,\frac{t}{m} \,.
\end{equation}
The comparison will be made clear in Fig. \ref{Fi:sql} to highlight the significance of the correlation effects.

At last we would like to make some remarks about the standard
quantum limit in some other often-studied examples in the context of
the measurement of the position of a quantum-mechanically free
mass~\cite{BA1}. In those examples, the basic idea is that the
evolution of the position of a free mass at time $t$ is given by
$q(t)=q(0)+p(0) t/m$. Thus, its position fluctuations with respect
to a prescribed quantum state then turn out to be
\begin{equation}
    \langle\Delta q^2(t)\rangle=\langle\Delta q^2(0)\rangle+\langle\Delta p^2(0)\rangle\,\frac{t^2}{m^2}+\langle\Delta q(0)\Delta p(0)+\Delta p(0)\Delta q(0)\rangle\,\frac{t}{m}\,.
\end{equation}
It is assumed that the correlation term in the last expression is
either vanishing or positive. The minimum position fluctuations is
achieved by considering the zero correlation. Then by applying the
minimum position-momentum uncertainty and minimizing the remaining expressions, we
obtain the SQL. However, in \cite{YU1}, Yuen points out that the
correlation term surely depends on the quantum state under
consideration, and can be negative in principle. Thus, in general,
beating the SQL is viable such that the position
fluctuations at later time can be made small with a proper choice of
the parameters, and even smaller than the SQL. The same conclusion
that the SQL may not be a lowest limit of the measurement
is shown in this mirror-field system.

\section{backreactions induced from radiation pressure}\label{sec4}
A stochastic description on dynamics of a mirror moving in quantum
fields has been studied in~\cite{WU}. There the associated
semiclassical Langevin equation is derived with the method of
Influence Functional by tracing out the quantum fields~\cite{WU}.
This Langevin equation incorporates not only the backreactions but
also noises manifested by the field fluctuations. To examine the effect of the backreaction, let us write down the Langevin equation up to first
order of the mirror's displacement,
\begin{equation}\label{E:EoM}
    m\ddot{q}(t)-i\int^{t}\!dt'\;\Theta(t-t')\,\langle\left[F_0(t),F_0(t')\right]\rangle_{\bar{\alpha}}\,q(t')-\langle F_0(t)\rangle_{\bar{\alpha}}-\langle\,\frac{\delta F}{\delta q}(t)\,\rangle_{\bar{\alpha}}\,q(t)=\xi_{\bar{\alpha}}\,.
\end{equation}
where $\Theta(\tau)$ is the unit-step function, and the expectation
value is taken with respect to the single-mode coherent state
$\lvert\bar{\alpha}\rangle$ of the scalar field. There are distinct contributions from the quantum scalar field. In addition to the mean radiation pressure given
 by
\begin{equation}
    \langle
    F_0(t)\rangle_{\bar{\alpha}}=\frac{1}{2}\int\!d\mathcal{A}_{\parallel}\;\Delta^{zz'}_{\bar{\alpha}}\bigl[z,t;z',t\bigr]_{z=z'=L}\,,
\end{equation}
there are two different backreaction effects~\cite{WU}. One is the
local variation of the mean radiation pressure due to the mean
displacement of the mirror,
\begin{equation}
 \langle\,\frac{\delta F}{\delta q} (t)\,\rangle_{\bar{\alpha}}\,q(t)=\frac{1}{2}\int\!d\mathcal{A}_{\parallel}\; \partial_L \,\Delta^{zz'}_{\bar{\alpha}}\bigl[z,t;z',t\bigr]_{z=z'=L} q(t)\,.
\end{equation}
The other comes from the nonlocal expression, associated with the retarded Green's function,
which is constructed from the expectation value of the commutator of
the forces. By the Wick's expansion, it can be written in terms of a
product of the Wightman functions $\Delta^{zz}_{\bar{\alpha}}$ and
$\chi_0$ of the scalar field in the case of the single-mode
coherent state,
\begin{equation}\label{FF}
    i\;\Theta( t-t')\,\langle\bigl[F(t),F (t')\bigr] \rangle_{\bar{\alpha}}
    =\Delta^{zz'}_{\bar{\alpha}}\bigl[z,t;z',t'\bigr]\,\chi_0 \bigl[z,t; z',t'\bigr]_{z=L+\epsilon,\,z'=L} \,,
\end{equation}
with
\begin{equation}
\chi_0 \bigl[z,t; z',t'\bigr] =i\,
\Theta(t-t')\int\!d\mathcal{A}_{\parallel}d\mathcal{A}'_{\parallel}\;\partial_z
\partial_{z'} \,\langle\bigl[\phi(x),\phi(x')\bigr]\rangle_{0}
 \,.
\end{equation}
Here we have used the point-splitting procedure to make this
expression well-defined, and will set the small separation parameter $\epsilon$
to zero in the end. In the coordinate representation, these two-point
functions take the form
\begin{align*}
   & \Delta^{zz'}_{\bar{\alpha}}(z,t;z',t')=\langle\partial_{z}\phi(x)\partial_{z}\phi(x')\rangle_{\bar{\alpha}}\notag\\
    &\quad\quad\quad=\frac{\lvert\bar{\alpha}\rvert^2\overline{\omega}}{2\pi^{3}}\,\cos\bigl[\overline{\omega}(z-L)\bigr]\cos\bigl[\overline{\omega}(z'-L)\bigr]\Big\{\cos\bigl[\overline{\omega}(t-t')\bigr]-\cos\bigl[\overline{\omega}(t+t'-2L)-2\varphi\bigr]\Big\}\\
   & i\,\langle\bigl[\phi(x),\phi(x')\bigr]\rangle_{0}=\frac{1}{4\pi R}\Bigl[\delta(t-t'-R)-\delta (t-t'-\bar{R})\Bigr]- \frac{1}{4\pi R}\Bigl[\delta(t-t'+R)-\delta (t-t'+\bar{R})\Bigr]\,,
\end{align*}
 respectively. Here $R$ and $\bar{R}$ are given by
\begin{equation*}
    R^2=(\mathbf{x}_{\parallel}-\mathbf{x'}_{\parallel})^2+(z-z')^2\,,\qquad\qquad\bar{R}^2=(\mathbf{x}_{\parallel}-\mathbf{x'}_{\parallel})^2+(z+z'-2L)^2\,,
\end{equation*}
and in particular $\bar{R}$ describes the distance between the field
point $(\mathbf{x}_{\parallel},z)$ and the image of the source point
$(\mathbf{x}_{\parallel},z')$ with respect to the mirror at $z=L$.

After plugging in the appropriate mode functions, the mean radiation
pressure and its variation are given by
\begin{eqnarray}
    \langle\,F(t)\,\rangle_{\bar{\alpha}}&=& \frac{\lvert\bar{\alpha}\rvert^2}{4\pi^{3}}\,\overline{\omega}\mathcal{A}_{\parallel}\left\{\vphantom{\Big|}1-\cos\bigl[2\overline{\omega}(t-L)-2\varphi\bigr]\right\}\,, \label{E:mwena}\\
     \langle\,\frac{\delta F}{\delta q} (t)\,\rangle_{\bar{\alpha}}&=& -\frac{\lvert\bar{\alpha}\rvert^2}{2\pi^{3}}\,\overline{\omega}^2\mathcal{A}_{\parallel}\left\{\vphantom{\Big|}\sin\bigl[2\overline{\omega}(t-L)-2\varphi\bigr]\right\}\, \label{E:vmwena}.
\end{eqnarray}
Although it is oscillatory, the mean radiation pressure of the
incoming coherent state is always positive. Thus
it keeps pushing the mirror along the positive $z$ axis. Similarly
we may carry out the integrals in Eq.~\eqref{FF}, and end up with,
\begin{align}\label{E:diss_bkrn}
    &\quad-i\int^{t}\!dt'\;\Theta(t-t')\,\langle\bigl[F(t),F(t')\bigr]\rangle_{\bar{\alpha}}\,q(t') \\
    &=-\frac{\lvert\bar{\alpha}\rvert^2\overline{\omega}}{4\pi^{4}}\,\cos\bigl[\overline{\omega}(z-L)\bigr]\cos\bigl[\overline{\omega}(z'-L)\bigr]\sin\bigl[\overline{\omega}(t-L)-\varphi\bigr]\int\!d\mathcal{A}_{\parallel}d\mathcal{A}'_{\parallel}\notag\\
    &\qquad\times\int^{t}\!dt'\;\Theta(t-t')\,\sin\bigl[\overline{\omega}(t'-L)-\varphi\bigr]\,q(t')\,\partial_{z}\partial_{z'}\left[\frac{1}{R}\,\delta(t-t'-R)-\frac{1}{\bar{R}}\,\delta(t-t'-\bar{R})\right]\,.\notag\\
    &=\frac{\lvert\bar{\alpha}\rvert^2}{2\pi^{3}}\,\overline{\omega}\mathcal{A}_{\parallel}\,\sin\bigl[\overline{\omega}(t-L)-\varphi\bigr]\,\Bigl\{\overline{\omega}\cos\bigl[\overline{\omega}(t-L)-\varphi\bigr]\,q(t)+\sin\bigl[\overline{\omega}(t-L)-\varphi\bigr]\,\dot{q}(t)\Bigr\}\,,\notag
\end{align}
where we have taken the limit $z=z'=L$ in the end. The outlines of
the calculations can be found in Appendix~\ref{E:sdjnlwpq}. This
nonlocal backreaction \eqref{E:diss_bkrn} reduces to local
expressions. Especially they include a term proportional to the
first-order time derivative of the displacement. Although the phase
of the incoming field changes with time, the coefficient of this
$\dot{q}(t)$ term remains always positive. Therefore it plays the
role of the frictional force, acting against the mirror's motion. It
is seen that  backreaction effects in
Eqs.~\eqref{E:vmwena}, ~\eqref{E:diss_bkrn} can be safely ignored as
long as $P\bar{\omega} t^2/m\ll1$ in the slow motion approximation,
$\bar{\omega}q(t) \ll 1$, as compared with the mean force term in
Eq.~\eqref{E:mwena}. Some remarks are in order. These backreaction
effects not only depend on the mirror's position and velocity but
also have time-dependent coefficients. In
particular, when the mirror is attached to an extra spring and undergoes
oscillatory motion~\cite{BO}, the presence of a time-dependent
coefficient of the position term may lead to instabilities on the mirror's
dynamics due to parametric oscillation. Then, it might
enhance the mirror's response of to small perturbations so as to
improve the sensitivity for measurement of the weak
signals~\cite{RU}. Therefore the dynamics of the mirror, when the
backreactions to the mirror are included, becomes rich.

The accompanying noise to the backreaction from the commutator of
the forces is denoted by $\xi_{\bar{\alpha}}$ in Eq.~\eqref{E:EoM},
which is
 associated with radiation pressure fluctuations,
\begin{equation}
    \xi_{\bar{\alpha}}=F_0(t)-\langle F_0(t)\rangle_{\bar{\alpha}}\,,
\end{equation}
as has been shown in Eq.~\eqref{q_F}. The consistent incorporation
of these two effects in the Langevin equation relies on the
existence of the relation between them, termed as the
fluctuations-dissipation relation. For our mirror-field system in
the case of single-mode coherent state, this relation can not be
presented in a transparent way. Let us write the noise-noise
autocorrelation function as
\begin{equation}\label{E:dflsjfn}
\frac{1}{2} \langle \, \bigl\{\xi (t), \, \xi(t')\bigr\} \,
\rangle_{\bar{\alpha}}=
\Delta^{zz'}_{\bar{\alpha}}\bigl[z,t;z',t'\bigr]\,\sigma_0
\bigl[z,t; z',t'\bigr]_{z=L+\epsilon,\,z'=L} \, ,
\end{equation}
where
\begin{equation}
\sigma_0 \bigl[z,t; z',t'\bigr] = \frac{1}{2}
\int\!d\mathcal{A}_{\parallel}d\mathcal{A}'_{\parallel}\;\partial_z
\partial_{z'} \,\langle\bigl\{ \phi(x),\phi(x')\bigr\} \rangle_{0}
 \,.
\end{equation}
is associated with anti-commutator of the scalar field in its vacuum state.
This noise-noise autocorrelation function \eqref{E:dflsjfn} can then be linked
to the retarded Green's function in~\eqref{FF} through a relation between the Fourier
transforms of the time-translationally invariant Green's functions
$\chi_{0}$ and $\sigma_{0}$. When evaluated at the mirror surface
$z=L$, $z'=L$, their Fourier transforms are respectively given by
 \begin{eqnarray}
 \chi_{0} \bigl[z,t; z',t'\bigr]_{z=L,\,z'=L} &=& \int \frac{d w}{2\pi} \, \chi_{0} ( \omega) \, e^{ -i
 \omega (t-t') } \, , \nonumber \\
 \sigma_{0} \bigl[z,t; z',t'\bigr]_{z=L,\,z'=L} &=& \int \frac{d w}{2\pi} \, \sigma_{0} ( \omega) \, e^{ -i
 \omega (t-t') } \, , \label{ft}
\end{eqnarray}
and the
fluctuations kernel $\sigma_{0} (\omega) $ can be shown to be related to the imaginary part of
the $ \chi_{0} ( \omega ) $ kernel,
\begin{equation}
\sigma_{0} ( \omega ) = {\rm Im} \left[ \chi_{0} ( \omega) \right]
\big[ \Theta (\omega)-\Theta (-\omega) \big] \,. \label{df-T}
\end{equation}
Thus we see that the role of the vacuum fluctuations of the field
seems to bridge the relation between the noise-noise autocorrelation
function and the Green's function constructed by the commutator of
the forces so that both effects can be treated in a consistent
manner in the Langevin equation.

The backreaction  effect has been proposed
theoretically in the work of Braginsky~\cite{BR} for its possible
role in setting the sensitivity limit of a detector in the Laser
Interferometer Gravitational Wave Observatory (LIGO). Recently,
radiation pressure cooling of a micromechanical oscillator has
been observed~\cite{KI}. It raise
a hope that quantum noise in a mirror-field system can be further
reduced by properly tuning the backreaction effect. In
Ref.~\cite{BU}, the backreaction has been extensively studied by
modifying the dynamics of the test mass within the ``optical spring"
scheme to improve performance of the laser interferometer
gravitational-wave detector. Here it is shown that these
backreactions can be naturally incorporated in the Langevin equation
by the field theoretic approach. Hence it will be a necessary
next step that we apply the Langevin equation \eqref{E:EoM} to the
interferometer in order to examine quantum noise as well as
backreactions.

\section{conclusions}\label{sec6}
The problem of quantum noise in the mirror-field system has been
studied in the field theoretic approach. We consider that a single,
perfectly reflecting mirror is illuminated by a single-mode coherent
state of a massless scalar field, in addition to ambient vacuum
fluctuations. The net field is read out by a monopole detector,
placed between the mirror and the field source. The radiation
pressure of the coherent state drives the mirror into motion. In the
slow motion limit, we may identify different sources of quantum
noise of the radiation field from readouts of the detector. In turn,
the effective distance between the mirror and the detector can be
obtained. The sources of quantum noise are respectively attributed
to the intrinsic fluctuations of the field (shot noise), induced
fluctuations arising from stochastic motion of the mirror due to the
radiation pressure fluctuations, and modified field fluctuations
which result from the mean displacement of the mirror. Their
correlations can then be established resulting from
interference between the incident field and the reflected field out
of the mirror in the read-out measurement. The overall uncertainty can be found
decreased (increased) due to negative (positive) correlation between
noise sources. In particular, negative correlation may lead to the
situation that overall uncertainty is even lower than that in the
standard quantum limit. Backreactions induced by the radiation
pressure is studied by deriving the associated Langevin equation
from first principles. The backreaction effects are found
insignificant for a slowly moving mirror. The Langevin equation
incorporates not only backreaction from radiation pressure on the
mirror but also noise manifested by the field fluctuations. Equipped
with the Langevin equation of the mirror-field system, it deserves
further study to improve the performance of the interferometer as
the work in~\cite{BU} by taking advantage of the backreaction
effect.

\appendix

\section{Variation of $I$}
Since we have shown the to the order we are interested, the ordering of operators are irrelevant, we will just write Eq.~\eqref{I_T_q2} as
\begin{equation}\label{E:skwhw}
    I=I_{0}+q\,\partial_L I_{0}+\frac{1}{2}\,q^2\,\partial^2_L I_{0}+\mathcal{O} (q^3)\,,
\end{equation}
where the subscript 0 denotes quantities evaluated on the mirror at $z=L$. We further define $\Delta O=O-\langle O\rangle$. Thus \eqref{E:skwhw} up to first order in $\Delta$ becomes
\begin{align}
    I&=\Bigl[\langle\,I_{0}\,\rangle+\Delta I_{0}\Bigr]+\Bigl[\Bigl(\langle\,q\,\rangle+\Delta q\Bigr)\Bigl(\langle\,\partial_L I_{0}\,\rangle+\Delta\partial_L I_{0}\Bigr)\Bigr]+\frac{1}{2}\Bigl[\Bigl(\langle\,q\,\rangle+\Delta q\Bigr)^{2}\Bigl(\langle\,\partial^{2}_L I_{0}\,\rangle+\Delta\partial^{2}_L I_{0}\Bigr)\Bigr]\notag\\
    &=\Bigl[\langle\,I_{0}\,\rangle+\Delta I_{0}\Bigr]+\Bigl[\langle\,q\,\rangle\langle\,\partial_LI_{0}\,\rangle+\Delta q\,\langle\,\partial_L I_{0}\,\rangle+\Delta\partial_L I_{0}\,\langle\,q\,\rangle\Bigr]\notag\\
    &\qquad\qquad\qquad\qquad+\frac{1}{2}\Bigl[\langle\,q\,\rangle^{2}\langle\,\partial^{2}_LI_{0}\,\rangle+2\Delta q\,\langle\,q\,\rangle\langle\,\partial^{2}_LI_{0}\,\rangle+\Delta\partial^{2}_L I_{0}\,\langle\,q\,\rangle^{2}\Bigr]+\mathcal{O}(\Delta^{2},\,q^{3})\,,\label{E:weoqsf}
\end{align}
Taking the expectation value yields
\begin{equation*}
    \langle\,I\,\rangle=\langle\,I_{0}\,\rangle+\langle\,q\,\rangle\langle\,\partial_LI_{0}\,\rangle+\frac{1}{2}\,\langle\,q\,\rangle^{2}\langle\,\partial^{2}_LI_{0}\,\rangle+\cdots\,,
\end{equation*}
and thus the quantity $\Delta I$ is given by
\begin{align}
    \Delta I&=\Delta I_{0}+\Bigl[\Delta q\,\langle\,\partial_L I_{0}\,\rangle+\Delta\partial_L I_{0}\,\langle\,q\,\rangle\Bigr]+\Bigl[\Delta q\,\langle\,q\,\rangle\langle\,\partial^{2}_LI_{0}\,\rangle+\frac{1}{2}\,\Delta\partial^{2}_L I_{0}\,\langle\,q\,\rangle^{2}\Bigr]+\cdots\,.
\end{align}
This is Eq.~\eqref{dz}. Since by definition of $\Delta O$, we have $\langle\,(\Delta O)^{2}\,\rangle=\langle\,\Delta^{2}O\,\rangle$; hence the variation of $I$ is
\begin{align}\label{E:woi}
    \langle\,\Delta^{2}I\,\rangle&=\langle\,\Delta^{2}I_{0}\,\rangle+\Bigl[\langle\,\Delta^{2}q\,\rangle\langle\,\partial_L I_{0}\,\rangle^{2}+2\langle\,\Delta q\,\Delta\partial_L I_{0}\,\rangle\langle\,q\,\rangle\langle\,\partial_L I_{0}\,\rangle+\langle\,\Delta^{2}(\partial_L I_{0})\,\rangle\langle\,q\,\rangle^{2}\Bigr]\notag\\
    &\qquad\qquad\qquad\qquad+\Bigl[2\langle\,\Delta I_{0}\,\Delta q\,\rangle\langle\,\partial_L I_{0}\,\rangle+2\langle\,\Delta I_{0}\,\Delta \partial_L I_{0}\,\rangle\langle\,q\,\rangle\Bigr]\\
    &\qquad\qquad\qquad\qquad\qquad\qquad+\Bigl[2\langle\,\Delta I_{0}\,\Delta q\,\rangle\langle\,q\,\rangle\langle\,\partial_L^{2}I_{0}\,\rangle+2\langle\,\Delta I_{0}\,\Delta \partial_L^{2}I_{0}\,\rangle\langle\,q\,\rangle^{2}\Bigr]+\cdots\,,\notag
\end{align}
where we have disregarded terms of the order $q^{2}\,\Delta q$ and higher because the motion of the mirror is minute in comparison with other length scales like $L$, $z_{0}$ and $\bar{\omega}^{-1}$.

\section{Evaluation of the subdominant terms in $\langle\Delta^{2}q\rangle_{\bar{\alpha}}$}\label{S:deltaq}
The displacement $q(t)$ of the mirror from its original position $z=L$ can be expressed as
\begin{equation}
    q(t)=\frac{1}{m}\int_{0}^{t}\!ds\!\int^{s}_{0}\!ds'\!\int\!d\mathcal{A}_{\parallel}\;:T_{zz}:(\mathbf{x}_{\parallel},z,s')\Big|_{z=L}\,,
\end{equation}
where the $zz$ component of the normal-ordered stress tensor $:T_{zz}:(\mathbf{x}_{\parallel},z,t)$ is given by
\begin{align*}
    :T_{zz}:(x_{\parallel},z,t)&=\frac{1}{2}\lim_{x'\to x}\partial_z\partial'_z\int'\!\frac{d^3k}{(2\pi)^{\frac{3}{2}}}\frac{1}{\sqrt{2\omega}}\int'\!\frac{d^3k'}{(2\pi)^{\frac{3}{2}}}\frac{1}{\sqrt{2\omega'}}\\
    &\qquad\times\Bigl\{\mathcal{U}_{\mathbf{k}}(\mathbf{x}_{\parallel},z,t)\mathcal{U}_{\mathbf{k}'}(\mathbf{x}'_{\parallel},z',t')\;a_ka_{k'}+\mathcal{U}^*_{\mathbf{k}}(\mathbf{x}_{\parallel},z,t)\mathcal{U}^*_{\mathbf{k}'}(\mathbf{x}'_{\parallel},z',t')\;a^{\dagger}_ka^{\dagger}_{k'}\Bigr.\\
    &\qquad\qquad\Bigl.+\;\mathcal{U}^*_{\mathbf{k}}(\mathbf{x}_{\parallel},z,t)\mathcal{U}_{\mathbf{k}'}(\mathbf{x}'_{\parallel},z',t')\;a^{\dagger}_ka^{\vphantom{\dagger}}_{k'}+\mathcal{U}_{\mathbf{k}}(\mathbf{x}_{\parallel},z,t)\mathcal{U}^*_{\mathbf{k}'}(\mathbf{x}'_{\parallel},z',t')\;a^{\dagger}_{k'}a_k^{\vphantom{\dagger}}\Bigr\}\,,
\end{align*}
in terms of the mode functions $\mathcal{U}_{\mathbf{k}}(\mathbf{x}_{\parallel},z,t)$. It is convenient to have
\begin{align*}
    \partial_{z}\mathcal{U}_{\mathbf{k}}(\mathbf{x}_{\parallel},z,t)\big|_{z=L}&=(2ik_{z})e^{i\mathbf{k}_{\parallel}\cdot\mathbf{x}_{\parallel}-i\omega t}\,e^{ik_zL}\,,\\
    \partial_{z}\mathcal{U}^{*}_{\mathbf{k}}(\mathbf{x}_{\parallel},z,t)\big|_{z=L}&=(-2ik_{z})e^{-i\mathbf{k}_{\parallel}\cdot\mathbf{x}_{\parallel}+i\omega t}\,e^{-ik_zL}\,,
\end{align*}
and define
\begin{align*}
    f_{\mathbf{k}\mathbf{k}'}(\mathbf{x}_{\parallel},t)&=\partial_{z}\mathcal{U}^{*}_{\mathbf{k}}(\mathbf{x}_{\parallel},z,t)\times\partial_{z'}\mathcal{U}_{\mathbf{k}'}(\mathbf{x}'_{\parallel},z',t')\big|^{x'=x}_{z=z'=L}\\
        &=4k_{z}k'_{z}\,e^{-i(\mathbf{k}_{\parallel}-\mathbf{k}'_{\parallel})\cdot\mathbf{x}_{\parallel}+i(\omega-\omega')t}e^{-i(k_{z}-k'_{z})L}\\
        &=f^{*}_{\mathbf{k}'\mathbf{k}}(\mathbf{x}_{\parallel},t)\,,\\
    g_{\mathbf{k}\mathbf{k}'}(\mathbf{x}_{\parallel},t)&=\partial_{z}\mathcal{U}_{\mathbf{k}}(\mathbf{x}_{\parallel},z,t)\times\partial_{z'}\mathcal{U}_{\mathbf{k}'}(\mathbf{x}'_{\parallel},z',t')\big|^{x'=x}_{z=z'=L}\\
        &=-4k_{z}k'_{z}\,e^{i(\mathbf{k}_{\parallel}+\mathbf{k}'_{\parallel})\cdot\mathbf{x}_{\parallel}-i(\omega+\omega')t}e^{i(k_{z}+k'_{z})L}\\
        &=g_{\mathbf{k}'\mathbf{k}}(\mathbf{x}_{\parallel},t)\,.
\end{align*}
The displacement operator $q(t)$ then takes a simpler form
\begin{equation*}
    q(t)=\frac{1}{2m}\int_{0}^{t}\!ds\!\int^{s}_{0}\!ds'\!\int\!\!d\mathcal{A}_{\parallel}\!\int'\!\!\frac{d^3k}{(2\pi)^{\frac{3}{2}}}\frac{1}{\sqrt{2\omega}}\!\int'\!\!\frac{d^3k'}{(2\pi)^{\frac{3}{2}}}\frac{1}{\sqrt{2\omega'}}\;\Bigl\{f_{\mathbf{k}\mathbf{k}'}(s')\,a^{\dagger}_{\mathbf{k}}a^{\vphantom{\dagger}}_{\mathbf{k}'}+g_{\mathbf{k}\mathbf{k}'}(s')\,a^{\vphantom{\dagger}}_{\mathbf{k}}a^{\vphantom{\dagger}}_{\mathbf{k}'}+\text{h.c.}\Bigr\}\,.
\end{equation*}
The calculations of $\langle\Delta^{2}q\rangle_{\bar{\alpha}}$ involve computations of the expressions like
\begin{itemize}
    \item $\langle a^{\dagger}_{\mathbf{k}}a^{\vphantom{\dagger}}_{\mathbf{k}'}a^{\dagger}_{\tilde{\mathbf{k}}}a^{\vphantom{\dagger}}_{\tilde{\mathbf{k}}'}\rangle_{\bar{\alpha}}-\langle a^{\dagger}_{\mathbf{k}}a^{\vphantom{\dagger}}_{\mathbf{k}'}\rangle_{\bar{\alpha}}\langle a^{\dagger}_{\tilde{\mathbf{k}}}a^{\vphantom{\dagger}}_{\tilde{\mathbf{k}}'}\rangle_{\bar{\alpha}}$,
    \item $\langle a^{\vphantom{\dagger}}_{\mathbf{k}}a^{\vphantom{\dagger}}_{\mathbf{k}'}a^{\dagger}_{\tilde{\mathbf{k}}}a^{\dagger}_{\tilde{\mathbf{k}}'}\rangle_{\bar{\alpha}}-\langle a^{\vphantom{\dagger}}_{\mathbf{k}}a^{\vphantom{\dagger}}_{\mathbf{k}'}\rangle_{\bar{\alpha}}\langle a^{\dagger}_{\tilde{\mathbf{k}}}a^{\dagger}_{\tilde{\mathbf{k}}'}\rangle_{\bar{\alpha}}$,
    \item $\langle a^{\dagger}_{\mathbf{k}}a^{\vphantom{\dagger}}_{\mathbf{k}'}a^{\dagger}_{\tilde{\mathbf{k}}}a^{\dagger}_{\tilde{\mathbf{k}}'}\rangle_{\bar{\alpha}}-\langle a^{\dagger}_{\mathbf{k}}a^{\vphantom{\dagger}}_{\mathbf{k}'}\rangle_{\bar{\alpha}}\langle a^{\dagger}_{\tilde{\mathbf{k}}}a^{\dagger}_{\tilde{\mathbf{k}}'}\rangle_{\bar{\alpha}}$, $\langle a^{\dagger}_{\mathbf{k}'}a^{\vphantom{\dagger}}_{\mathbf{k}}a^{\dagger}_{\tilde{\mathbf{k}}}a^{\dagger}_{\tilde{\mathbf{k}}'}\rangle_{\bar{\alpha}}-\langle a^{\dagger}_{\mathbf{k}'}a^{\vphantom{\dagger}}_{\mathbf{k}}\rangle_{\bar{\alpha}}\langle a^{\dagger}_{\tilde{\mathbf{k}}}a^{\dagger}_{\tilde{\mathbf{k}}'}\rangle_{\bar{\alpha}}$, and their complex conjugates,
\end{itemize}
and so on. The result of the first term is given by Eq.~\eqref{rpfa+a}. Here we would like to evaluate the second and the third terms, and compare them with the first term.

\subsection{Evaluation of second term}
Since
\begin{align*}
    \langle a^{\vphantom{\dagger}}_{\mathbf{k}}a^{\vphantom{\dagger}}_{\mathbf{k}'}a^{\dagger}_{\tilde{\mathbf{k}}}a^{\dagger}_{\tilde{\mathbf{k}}'}\rangle_{\bar{\alpha}}-\langle a^{\vphantom{\dagger}}_{\mathbf{k}}a^{\vphantom{\dagger}}_{\mathbf{k}'}\rangle_{\bar{\alpha}}\langle a^{\dagger}_{\tilde{\mathbf{k}}}a^{\dagger}_{\tilde{\mathbf{k}}'}\rangle_{\bar{\alpha}}&=\lvert\bar{\alpha}\rvert^{2}\Bigl[\delta(\tilde{\mathbf{k}}-\bar{\mathbf{k}})\delta(\mathbf{k}'-\bar{\mathbf{k}})\delta(\mathbf{k}-\tilde{\mathbf{k}}')+\delta(\mathbf{k}'-\bar{\mathbf{k}})\delta(\tilde{\mathbf{k}}'-\bar{\mathbf{k}})\delta(\mathbf{k}-\tilde{\mathbf{k}})\Bigr.\\
    &\qquad\Bigl.+\;\delta(\mathbf{k}-\bar{\mathbf{k}})\delta(\tilde{\mathbf{k}}-\bar{\mathbf{k}})\delta(\mathbf{k}'-\tilde{\mathbf{k}}')+\delta(\mathbf{k}-\bar{\mathbf{k}})\delta(\tilde{\mathbf{k}}'-\bar{\mathbf{k}})\delta(\mathbf{k}'-\tilde{\mathbf{k}})\Bigr]\\
    &\qquad\qquad+\delta(\mathbf{k}-\tilde{\mathbf{k}})\delta(\mathbf{k}'-\tilde{\mathbf{k}}')+\delta(\mathbf{k}-\tilde{\mathbf{k}}')\delta(\mathbf{k}'-\tilde{\mathbf{k}})\,,
\end{align*}
and suppose that the particle number density $\lvert\bar{\alpha}\rvert^{2}$ is sufficiently large such that the pure vacuum contribution can be ignored, the contribution from $\langle a^{\vphantom{\dagger}}_{\mathbf{k}}a^{\vphantom{\dagger}}_{\mathbf{k}'}a^{\dagger}_{\tilde{\mathbf{k}}}a^{\dagger}_{\tilde{\mathbf{k}}'}\rangle_{\bar{\alpha}}-\langle a^{\vphantom{\dagger}}_{\mathbf{k}}a^{\vphantom{\dagger}}_{\mathbf{k}'}\rangle_{\bar{\alpha}}\langle a^{\dagger}_{\tilde{\mathbf{k}}}a^{\dagger}_{\tilde{\mathbf{k}}'}\rangle_{\bar{\alpha}}$ to the variance $\langle\,\Delta^{2}q(t)\,\rangle_{\bar{\alpha}}$ is then given by
\begin{align}
    &\qquad\frac{\lvert\bar{\alpha}\rvert^{2}}{m^{2}}\!\int_{0}^{t}\!ds\!\int^{s}_{0}\!ds'\!\int\!\!d\mathcal{A}_{\parallel}\!\int_{0}^{t}\!d\tau\!\int^{\tau}_{0}\!d\tau'\!\int\!\!d\mathcal{A}'_{\parallel}\!\int'\!\!\frac{d^3k}{(2\pi)^{3}}\frac{1}{2\omega}\frac{1}{(2\pi)^{3}}\frac{1}{2\bar{\omega}}\;g_{\mathbf{k}\bar{\mathbf{k}}}(\mathbf{x}_{\parallel},s')g^{*}_{\mathbf{k}\bar{\mathbf{k}}}(\mathbf{x}'_{\parallel},\tau')\notag\\
    &=16\,\frac{\lvert\bar{\alpha}\rvert^{2}}{m^{2}}\frac{\mathcal{A}_{\parallel}}{(2\pi)^{3}}\frac{\bar{\omega}}{2}\!\int_{0}^{\infty}\!\frac{d\omega}{2\pi}\;\frac{\omega}{2}\!\int_{0}^{t}\!ds\!\int^{s}_{0}\!ds'\!\int_{0}^{t}\!d\tau\!\int^{\tau}_{0}\!d\tau'\;e^{-i(\omega+\bar{\omega})s'}e^{+i(\omega+\bar{\omega})\tau'}\,,\label{E:dksndksf}
\end{align}
where we have assumed that the incoming coherent field propagates along the $z$ direction, i.e., $\bar{\mathbf{k}}_{\parallel}=0$. After integrating the time variables first, we are left with an $\omega$-integral, which is formally UV-divergent. We regularize it with a cutoff frequency $\Lambda$, and then the result of \eqref{E:dksndksf} is
\begin{equation*}
    2\,\frac{\lvert\bar{\alpha}\rvert^{2}\bar{\omega}}{m^{2}\pi}\frac{\mathcal{A}_{\parallel}t^{2}}{(2\pi)^{3}}\left\{\left[\ln\frac{\Lambda+\bar{\omega}}{\bar{\omega}}-\left(1+\frac{\bar{\omega}}{\Lambda}\right)^{-1}\right]+\frac{1}{3\bar{\omega}^{2}t^{2}}+\mathcal{O}(\frac{1}{\bar{\omega}^{3}t^{3}})\right\}\,,
\end{equation*}
in the long time limit $\bar{\omega}t\gg1$. The cutoff-dependent term is logarithmic. For any sensible value of $\Lambda$, it should be of the order $\mathcal{O}(1)$. Thus the expressions in the curly brackets are at most of the order unity. Compared with \eqref{rpfa+a}, this contribution is of about the order $\mathcal{O}(\bar{\omega}^{-1}t^{-1})$ smaller; thus it is relatively subdominant. Note that the contribution from the expression $\langle a^{\dagger}_{\mathbf{k}}a^{\dagger}_{\mathbf{k}'}a^{\vphantom{\dagger}}_{\tilde{\mathbf{k}}}a^{\vphantom{\dagger}}_{\tilde{\mathbf{k}}'}\rangle_{\bar{\alpha}}-\langle a^{\dagger}_{\mathbf{k}}a^{\dagger}_{\mathbf{k}'}\rangle_{\bar{\alpha}}\langle a^{\vphantom{\dagger}}_{\tilde{\mathbf{k}}}a^{\vphantom{\dagger}}_{\tilde{\mathbf{k}}'}\rangle_{\bar{\alpha}}$ is identically zero for a single-mode coherent field.

\subsection{Evaluation of the third therm}
From
\begin{align*}
    \langle a^{\dagger}_{\mathbf{k}}a^{\vphantom{\dagger}}_{\mathbf{k}'}a^{\dagger}_{\tilde{\mathbf{k}}}a^{\dagger}_{\tilde{\mathbf{k}}'}\rangle_{\bar{\alpha}}-\langle a^{\dagger}_{\mathbf{k}}a^{\vphantom{\dagger}}_{\mathbf{k}'}\rangle_{\bar{\alpha}}\langle a^{\dagger}_{\tilde{\mathbf{k}}}a^{\dagger}_{\tilde{\mathbf{k}}'}\rangle_{\bar{\alpha}}&=\bar{\alpha}^{*2}\Bigl[\delta(\mathbf{k}-\bar{\mathbf{k}})\delta(\tilde{\mathbf{k}}-\bar{\mathbf{k}})\delta(\mathbf{k}'-\tilde{\mathbf{k}}')+\delta(\mathbf{k}-\bar{\mathbf{k}})\delta(\tilde{\mathbf{k}}'-\bar{\mathbf{k}})\delta(\mathbf{k}'-\tilde{\mathbf{k}})\Bigr]
\end{align*}
with $\bar{\alpha}^{*}$ being a complex conjugate of the complex coherent parameter $\bar{\alpha}$, its the contribution to the variance $\langle\,\Delta^{2}q(t)\,\rangle_{\bar{\alpha}}$ is
\begin{align}
    &\qquad\frac{\bar{\alpha}^{*2}}{2m^{2}}\!\int_{0}^{t}\!ds\!\int^{s}_{0}\!ds'\!\int\!\!d\mathcal{A}_{\parallel}\!\int_{0}^{t}\!d\tau\!\int^{\tau}_{0}\!d\tau'\!\int\!\!d\mathcal{A}'_{\parallel}\!\int'\!\!\frac{d^3k}{(2\pi)^{3}}\frac{1}{2\omega}\frac{1}{(2\pi)^{3}}\frac{1}{2\bar{\omega}}\;f_{\bar{\mathbf{k}}\mathbf{k}}(\mathbf{x}_{\parallel},s')g^{*}_{\bar{\mathbf{k}}\mathbf{k}}(\mathbf{x}'_{\parallel},\tau')\notag\\
    &=-8\,\frac{\bar{\alpha}^{*2}}{m^{2}}\frac{\mathcal{A}_{\parallel}}{(2\pi)^{3}}\frac{\bar{\omega}}{2}\,e^{-i\,2\bar{\omega}L}\!\int_{0}^{\infty}\!\frac{d\omega}{2\pi}\;\frac{\omega}{2}\!\int_{0}^{t}\!ds\!\int^{s}_{0}\!ds'\!\int_{0}^{t}\!d\tau\!\int^{\tau}_{0}\!d\tau'\;e^{-i(\omega-\bar{\omega})s'}e^{+i(\omega+\bar{\omega})\tau'}\,.\label{E:sndksf}
\end{align}
We use the following approximations in the long time limit to simplify the $\omega$-integral,
\begin{align*}
    \int_{0}^{t}\!ds\!\int^{s}_{0}\!ds'\;e^{-i(\omega-\bar{\omega})s'}&=2\,\frac{\sin^{2}(\omega-\bar{\omega})\frac{t}{2}}{(\omega-\bar{\omega})^{2}}-i\,\left[\frac{t}{\omega-\bar{\omega}}-\frac{\sin(\omega-\bar{\omega})t}{(\omega-\bar{\omega})^{2}}\right]\\
        &\approx2\pi\,\frac{\sin(\omega-\bar{\omega})\frac{t}{2}}{(\omega-\bar{\omega})}\,\delta(\omega-\bar{\omega})-i\,\left[\frac{t}{\omega-\bar{\omega}}-\frac{\sin(\omega-\bar{\omega})t}{(\omega-\bar{\omega})^{2}}\right]\\
    \int_{0}^{t}\!d\tau\!\int^{\tau}_{0}\!d\tau'\;e^{+i(\omega+\bar{\omega})\tau'}&=2\,\frac{\sin^{2}(\omega+\bar{\omega})\frac{t}{2}}{(\omega+\bar{\omega})^{2}}+i\,\left[\frac{t}{\omega+\bar{\omega}}-\frac{\sin(\omega+\bar{\omega})t}{(\omega+\bar{\omega})^{2}}\right]\\
        &\approx i\,\left[\frac{t}{\omega+\bar{\omega}}-\frac{\sin(\omega+\bar{\omega})t}{(\omega+\bar{\omega})^{2}}\right]\,,
\end{align*}
due to the fact that $\bar{\omega}t\gg1$ and $\bar{\omega}>0$. Similar to the previous case, the integral over $\omega$ is divergent, so a cutoff is introduced to regularize the result and Eq.~\eqref{E:sndksf} becomes
\begin{equation*}
    -2\,\frac{\bar{\alpha}^{*2}\bar{\omega}}{m^{2}\pi}\frac{\mathcal{A}_{\parallel}t^{2}}{(2\pi)^{3}}\,e^{-i\,2\bar{\omega}L}\left[\ln\frac{\Lambda}{\bar{\omega}}+i\,\frac{\pi}{4}+\mathcal{O}(\frac{1}{\bar{\omega}t})\right]\,.
\end{equation*}
There are three other similar terms and their results are either the same as this one or its complex conjugate. Following the arguments in the previous subsection, contributions from this type of expressions are negligible in comparison with \eqref{rpfa+a} in the large time limit. We also note the contribution from $\langle a^{\dagger}_{\mathbf{k}}a^{\dagger}_{\mathbf{k}'}a^{\dagger}_{\tilde{\mathbf{k}}}a^{\vphantom{\dagger}}_{\tilde{\mathbf{k}}'}\rangle_{\bar{\alpha}}-\langle a^{\dagger}_{\mathbf{k}}a^{\dagger}_{\mathbf{k}'}\rangle_{\bar{\alpha}}\langle a^{\dagger}_{\tilde{\mathbf{k}}}a^{\vphantom{\dagger}}_{\tilde{\mathbf{k}}'}\rangle_{\bar{\alpha}}$ vanishes identically.

As the final remarks, there are no contributions from $\langle a^{\dagger}_{\mathbf{k}}a^{\dagger}_{\mathbf{k}'}a^{\dagger}_{\tilde{\mathbf{k}}}a^{\dagger}_{\tilde{\mathbf{k}}'}\rangle_{\bar{\alpha}}-\langle a^{\dagger}_{\mathbf{k}}a^{\dagger}_{\mathbf{k}'}\rangle_{\bar{\alpha}}\langle a^{\dagger}_{\tilde{\mathbf{k}}}a^{\dagger}_{\tilde{\mathbf{k}}'}\rangle_{\bar{\alpha}}$ and $\langle a^{\vphantom{\dagger}}_{\mathbf{k}}a^{\vphantom{\dagger}}_{\mathbf{k}'}a^{\vphantom{\dagger}}_{\tilde{\mathbf{k}}}a^{\vphantom{\dagger}}_{\tilde{\mathbf{k}}'}\rangle_{\bar{\alpha}}-\langle a^{\vphantom{\dagger}}_{\mathbf{k}}a^{\vphantom{\dagger}}_{\mathbf{k}'}\rangle_{\bar{\alpha}}\langle a^{\vphantom{\dagger}}_{\tilde{\mathbf{k}}}a^{\vphantom{\dagger}}_{\tilde{\mathbf{k}}'}\rangle_{\bar{\alpha}}$ because they vanishes identically, too.

\section{Evaluation of Eq.~\eqref{E:diss_bkrn}}\label{E:sdjnlwpq}
We are interested in evaluating part of \eqref{E:diss_bkrn}, denoted by $\mathcal{J}$,
\begin{equation*}
    \mathcal{J}=\int\!d\mathcal{A}_{\parallel}d\mathcal{A}'_{\parallel}\int\!dt'\;\Theta(t-t')\sin[\bar{\omega}(t'-L)-\varphi]\Bigl(\partial_{z}\partial_{z'}\left[\phi(x),\phi(x')\right]\Bigr)\delta q(t')\,.
\end{equation*}
The retarded Green's function can be decomposed into two components; one corresponds to unbounded Minkowski space and the other to the image contribution with respect to mirror's original position $z=L$. We consider the unbounded component for the moment, and leave the image contribution to later sections.

Let us take the partial differentiations $\partial_{z} $ before integrals over the cross sectional area $\mathcal{A}_{\parallel}$. Here we will assume the the linear dimension $\sqrt{\mathcal{A}_{\parallel}}$ is much larger than $\overline{\omega}^{-1}$. Thus the evaluation of $\mathcal{J}$ may proceed as
\begin{align*}
    \mathcal{J}&=\int\!d\mathcal{A}_{\parallel}d\mathcal{A}'_{\parallel}\,\partial_{z}\partial_{z'}\!\int\!dt'\;\sin[\bar{\omega}(t'-L)-\varphi]\theta(t-t')\,\frac{1}{R}\,\delta(t-t'-R)\,q(t')\\
    &=\int\!d\mathcal{A}_{\parallel}d\mathcal{A}'_{\parallel}\;\partial_{z}\partial_{z'}\left\{\sin(\vartheta-\bar{\omega}R)\,\frac{1}{R}\,q(t-R)\right\}\,,\qquad\qquad\qquad\vartheta=\bar{\omega}(t-L)-\varphi\,,\\
    &=\sum_{n=0}^{\infty}\frac{(-1)^{n}}{n!}\,q^{(n)}(t)\int\!d\mathcal{A}_{\parallel}d\mathcal{A}'_{\parallel}\;\partial_{z}\partial_{z'}\Bigl\{\sin(\vartheta-\bar{\omega}R)\,R^{n-1}\Bigr\}\\
    &=2\pi\mathcal{A}_{\parallel}\sum_{n=0}^{\infty}\frac{(-1)^{n}}{n!}\,q^{(n)}(t)\int_{\lvert\epsilon\rvert}^{\infty}\!dR\;\biggl\{\bar{\omega}^{2}\epsilon^{2}R^{n-2}\sin(\vartheta-\bar{\omega}R)+\bar{\omega}\left(R^{2}-\epsilon^{2}\right)R^{n-3}\cos(\vartheta-\bar{\omega}R)\biggr.\notag\\
    &\qquad\qquad\qquad+2(n-1)\bar{\omega}\epsilon^{2}R^{n-3}\cos(\vartheta-\bar{\omega}R)-(n-1)\left(R^{2}-\epsilon^{2}\right)R^{n-4}\sin(\vartheta-\bar{\omega}R)\notag\\
    &\qquad\qquad\qquad\qquad\qquad\qquad\qquad\qquad\biggl.-\,(n-1)(n-2)\epsilon^{2}R^{n-4}\sin(\vartheta-\bar{\omega}R)\biggr\}\,,
\end{align*}
where we have assume that the Taylor's expansion of $q(t-R)$ with respect to $R$ exists, and that implies the assumption that the memory of the system is rather short. We also have used the identities,
\begin{align*}
    \frac{\partial R}{\partial z}&=-\frac{\partial R}{\partial z'}=\frac{z-z'}{R}=\frac{\epsilon}{R}\,,\\
    \frac{\partial^{2}R}{\partial z^{2}}&=\frac{\partial^{2}R}{\partial z'^{2}}=\frac{1}{R}-\frac{(z-z')^{2}}{R^{3}}=\frac{1}{R}-\frac{\epsilon^{2}}{R^{3}}\,,
\end{align*}
with $\epsilon=z-z'\lessgtr0$. For later convenience we will set $\varepsilon=\lvert\epsilon\rvert$, and let $\varepsilon\to 0$ in the end of the calculations.

First, we check $n=2$ case, and the integral over $R$ becomes
\begin{align*}
    \mathcal{K}_{2}&=\int_{\varepsilon}^{\infty}\!dR\;\biggl\{\bar{\omega}^{2}\varepsilon^{2}\sin(\vartheta-\bar{\omega}R)+\bar{\omega}\left(R-\frac{\varepsilon^{2}}{R}\right)\cos(\vartheta-\bar{\omega}R)\biggr.\notag\\
    &\qquad\qquad\qquad\qquad\qquad\qquad\qquad\qquad\biggl.+\,2\bar{\omega}\,\frac{\varepsilon^{2}}{R}\cos(\vartheta-\bar{\omega}R)-\left(1-\frac{\varepsilon^{2}}{R^{2}}\right)\sin(\vartheta-\bar{\omega}R)\biggr\}\\
    &=\int_{\varepsilon}^{\infty}\!dR\;\biggl\{\Bigl[\bar{\omega}R\cos(\vartheta-\bar{\omega}R)-\sin(\vartheta-\bar{\omega}R)\Bigr]+\frac{\varepsilon^{2}}{R^{2}}\Bigl[\bar{\omega}R\cos(\vartheta-\bar{\omega}R)+\sin(\vartheta-\bar{\omega}R)\Bigr]\biggr.\notag\\
    &\qquad\qquad\qquad\qquad\biggl.+\,\bar{\omega}^{2}\varepsilon^{2}\sin(\vartheta-\bar{\omega}R)\biggr\}\,,
\end{align*}
where
\begin{align*}
    \int_{\varepsilon}^{\infty}\!dR\;\frac{\varepsilon^{2}}{R^{2}}\Bigl[\bar{\omega}R\cos(\vartheta-\bar{\omega}R)+\sin(\vartheta-\bar{\omega}R)\Bigr]&=\varepsilon\sin(\vartheta-\bar{\omega}\varepsilon)\,,\\
    \int_{\varepsilon}^{\infty}\!dR\;\Bigl[\bar{\omega}R\cos(\vartheta-\bar{\omega}R)-\sin(\vartheta-\bar{\omega}R)\Bigr]&=\varepsilon\sin(\vartheta-\bar{\omega}\varepsilon)\,,\\
    \int_{\varepsilon}^{\infty}\!dR\;\bar{\omega}^{2}\varepsilon^{2}\sin(\vartheta-\bar{\omega}R)&=-\bar{\omega}^{2}\varepsilon\cos(\vartheta-\bar{\omega}\varepsilon)\,.
\end{align*}
Thus the integral $\mathcal{K}_{2}$ is of the order $\mathcal{O}(\varepsilon)$ for the $n=2$ case. Note that some of the integrals are still not well-defined in the large $R$ regime so we introduce some cutoff to regularize them, but it turns out that the results are independent of the ultraviolet cutoff.

Similarly for $n=1$, we have
\begin{align*}
    \mathcal{K}_{1}&=\int_{\varepsilon}^{\infty}\!dR\;\biggl\{\bar{\omega}^{2}\frac{\varepsilon^{2}}{R}\sin(\vartheta-\bar{\omega}R)+\bar{\omega}\left(1-\frac{\varepsilon^{2}}{R^{2}}\right)\cos(\vartheta-\bar{\omega}R)\biggr\}\\
    &=\int_{\varepsilon}^{\infty}\!dR\;\biggl\{\bar{\omega}\cos(\vartheta-\bar{\omega}R)+\frac{\bar{\omega}\varepsilon^{2}}{R^{2}}\Bigl[\bar{\omega}R\sin(\vartheta-\bar{\omega}R)-\cos(\vartheta-\bar{\omega}R)\Bigr]\biggr\}\\
    &=\int_{\varepsilon}^{\infty}\!dR\;\bar{\omega}\cos(\vartheta-\bar{\omega}R)+\mathcal{O}(\varepsilon)\\
    &=\sin\vartheta+\mathcal{O}(\varepsilon)\,,
\end{align*}
where
\begin{equation*}
    \int_{\varepsilon}^{\infty}\!dR\;\frac{\bar{\omega}\varepsilon^{2}}{R^{2}}\Bigl[\bar{\omega}R\sin(\vartheta-\bar{\omega}R)-\cos(\vartheta-\bar{\omega}R)\Bigr]=-\bar{\omega}\varepsilon\cos(\vartheta-\bar{\omega}\varepsilon)\,.
\end{equation*}

Finally in the case $n=0$, the integral $\mathcal{K}_{0}$ is given by
\begin{align*}
    \mathcal{K}_{0}&=\int_{\lvert\epsilon\rvert}^{\infty}\!dR\;\biggl\{\frac{\bar{\omega}^{2}\epsilon^{2}}{R^{2}}\sin(\vartheta-\bar{\omega}R)+\bar{\omega}\left(\frac{1}{R}-\frac{\epsilon^{2}}{R^{3}}\right)\cos(\vartheta-\bar{\omega}R)-2\frac{\epsilon^{2}}{R^{4}}\sin(\vartheta-\bar{\omega}R)\biggr.\notag\\
    &\biggl.\qquad\qquad\qquad\qquad\qquad\qquad-\,2\frac{\bar{\omega}\epsilon^{2}}{R^{3}}\cos(\vartheta-\bar{\omega}R)+\left(\frac{1}{R^{2}}-\frac{\epsilon^{2}}{R^{4}}\right)\sin(\vartheta-\bar{\omega}R)\biggr\}\notag\\
    &=-\bar{\omega}\cos(\vartheta-\bar{\omega}\varepsilon)=-\bar{\omega}\cos\vartheta+\mathcal{O}(\varepsilon)\,.
\end{align*}
In general, for $n\geq2$, it can be shown that
\begin{equation*}
    \mathcal{K}_{n}=\varepsilon^{n-1}\Bigl[-\bar{\omega}\varepsilon\cos(\vartheta-\bar{\omega}\varepsilon)+n\sin(\vartheta-\bar{\omega}\varepsilon)\Bigr]\,,
\end{equation*}
so $\mathcal{K}_{n}$ behaves like $\mathcal{O}(\varepsilon^{n-1})$ for $n\geq2$. Therefore $\mathcal{J}$ is
\begin{equation}
    \mathcal{J}=-2\pi\mathcal{A}_{\parallel}\Bigl\{\bar{\omega}\cos\vartheta\,\delta q(t)+\sin\vartheta\,\delta\dot{q}(t)+\mathcal{O}(\varepsilon)\Bigr\}\,.
\end{equation}

\begin{acknowledgments}
DSL would like to thank W. G. Unruh, B. L. Hu and L. H. Ford for stimulating discussions and useful suggestions. This work was supported in part by the National Science Council, R. O. C. under grant NSC97-2112-M-259-007-MY3, and the National Center for Theoretical Sciences, Taiwan.
\end{acknowledgments}

\end{document}